\documentclass[12pt]{article}
\usepackage{amsfonts,amsmath,amssymb,epsf}
\usepackage{amsmath,braket}
\usepackage{physics}
\usepackage{graphicx,hyperref,color}
\hypersetup{
    bookmarks=true,         
    unicode=false,          
    pdftoolbar=true,        
    pdfmenubar=true,        
    linktocpage=true,       
    pdffitwindow=false,     
    pdfstartview={FitH},    
    pdfnewwindow=true,      
    colorlinks=true,       
    linkcolor=blue,          
    citecolor=blue,        
    filecolor=blue,      
    urlcolor=blue           
}

\numberwithin{equation}{section}									

\newcommand{\de}{\partial}
\newcommand{\be}{\begin{equation}}
\newcommand{\ba}{\begin{eqnarray}}
\newcommand{\ea}{\end{eqnarray}}
\newcommand{\ee}{\end{equation}}

\newcommand{\f}{\frac}
\newcommand{\s}{\sqrt}
\newcommand{\vp}{\varphi}

\newcommand{\ti}{\tilde}
\newcommand{\ap}{\alpha}

\newcommand{\ddd}{\cdot\cdot\cdot}
\newcommand{\no}{\nonumber \\}
\newcommand{\la}{\langle}
\newcommand{\lb}{\rangle}
\newcommand{\bea}{\begin{eqnarray}}
\newcommand{\eea}{\end{eqnarray}}
\newcommand{\bes}{\begin{equation*}}
\newcommand{\beas}{\begin{eqnarray*}}
\newcommand{\eeas}{\end{eqnarray*}}
\newcommand{\bas}{\begin{array*}}
\newcommand{\eas}{\end{array*}}
\newcommand{\ees}{\end{equation*}}

\newcommand{\ep}{\epsilon}

   \let\vp=\varphi

\begin{document}

\begin{titlepage}
\thispagestyle{empty}

\vspace*{-2cm}
\begin{flushright}
YITP-22-71
\\
IPMU22-0036
\\
\end{flushright}

\bigskip

\begin{center}
\noindent{{\large \textbf{Wedge Holography in Flat Space and Celestial Holography}}}\\
\vspace{2cm}

Naoki Ogawa$^{a}$,\ Tadashi Takayanagi$^{a,b,c}$,\ Takashi Tsuda$^a$, and Takahiro Waki$^a$
\vspace{1cm}\\

{\it $^a$Center for Gravitational Physics and Quantum Information,\\
Yukawa Institute for Theoretical Physics, Kyoto University, \\
Kitashirakawa Oiwakecho, Sakyo-ku, Kyoto 606-8502, Japan}\\
\vspace{1mm}
{\it $^b$Inamori Research Institute for Science,\\
620 Suiginya-cho, Shimogyo-ku,
Kyoto 600-8411 Japan}\\
\vspace{1mm}
{\it $^{c}$Kavli Institute for the Physics and Mathematics
 of the Universe (WPI),\\
University of Tokyo, Kashiwa, Chiba 277-8582, Japan}\\

\bigskip \bigskip
\vskip 2em
\end{center}

\begin{abstract}
In this paper, we study codimension two holography in flat spacetimes, based on the idea of the wedge holography. We propose that a region in a $d+1$ dimensional flat spacetime surrounded by two end of the world-branes, which are given by $d$ dimensional hyperbolic spaces, is dual to a conformal field theory (CFT) on a $d-1$ dimensional sphere. Similarly, we also propose that a $d+1$ dimensional region in the flat spacetime bounded by two $d$ dimensional de Sitter spaces is holographically dual to a CFT on a $d-1$ dimensional sphere. Our calculations of the partition function, holographic entanglement entropy and two point functions, support these duality relations and imply that such CFTs are non-unitary. Finally, we glue these two dualities along null surfaces to realize a codimension two holography for a full Minkowski spacetime and discuss a possible connection to the celestial holography.

\end{abstract}

\end{titlepage}

\newpage

\tableofcontents

\section{Introduction}
\label{sec:intro}

The holographic principle \cite{tHooft:1993dmi,Susskind:1994vu} usually relates a gravitational theory on a certain spacetime $M$ to a non-gravitational theory on its codimension one boundary $\de M$. This holographic property is manifest in the AdS/CFT \cite{Maldacena:1997re} and the dS/CFT \cite{Strominger:2001pn,Maldacena:2002vr}.
However, if we try to apply the usual analysis of bulk to boundary relation in the AdS/CFT \cite{Gubser:1998bc,Witten:1998qj} to a $d+1$ dimensional flat Lorentzian spacetime, its mathematical structure strongly implies that the dual theory is a $d-1$ dimensional conformal field theory (CFT) which lives on a sphere at null infinity \cite{deBoer:2003vf}. Motivated by the triangle equivalence between the soft theorems, memory effects and BMS symmetries \cite{Strominger:2013jfa,He:2014laa,He:2014cra,Strominger:2014pwa,Pasterski:2015zua,Strominger:2017zoo,Capone:2022gme}, the celestial holography \cite{He:2015zea,Pasterski:2016qvg,Pasterski:2017kqt,Pasterski:2017ylz,Bagchi:2016bcd,Cardona:2017keg,Cheung:2016iub} was proposed.\footnote{
A similar codimension two holography was argued in \cite{Freivogel:2006xu} in the context of eternal inflation. Refer to e.g, \cite{Dappiaggi:2004kv,Donnay:2022aba} for a proposal of holographic duality between gravity in four dimensional Minkowski spacetime and a three dimensional conformal Carrollian field theory. Also see \cite{Li:2010dr} for a possibility of a codimension one holography between gravity in the $d+1$ dimensional Euclidean flat space R$^{d+1}$ and  a $d$ dimensional CFT on S$^{d}$.} This interesting holographic duality argues that the four dimensional gravity on an asymptotically flat spacetime is equivalent to a two dimensional CFT at null infinity, such that the S-matrices of the four dimensional gravity can be computed from correlation functions in the two dimensional CFT via a certain Mellin-like transformation, though the precise identification of the dual CFT has remained to be answered.

\begin{figure}[hhh]
  \centering
  \includegraphics[width=8cm]{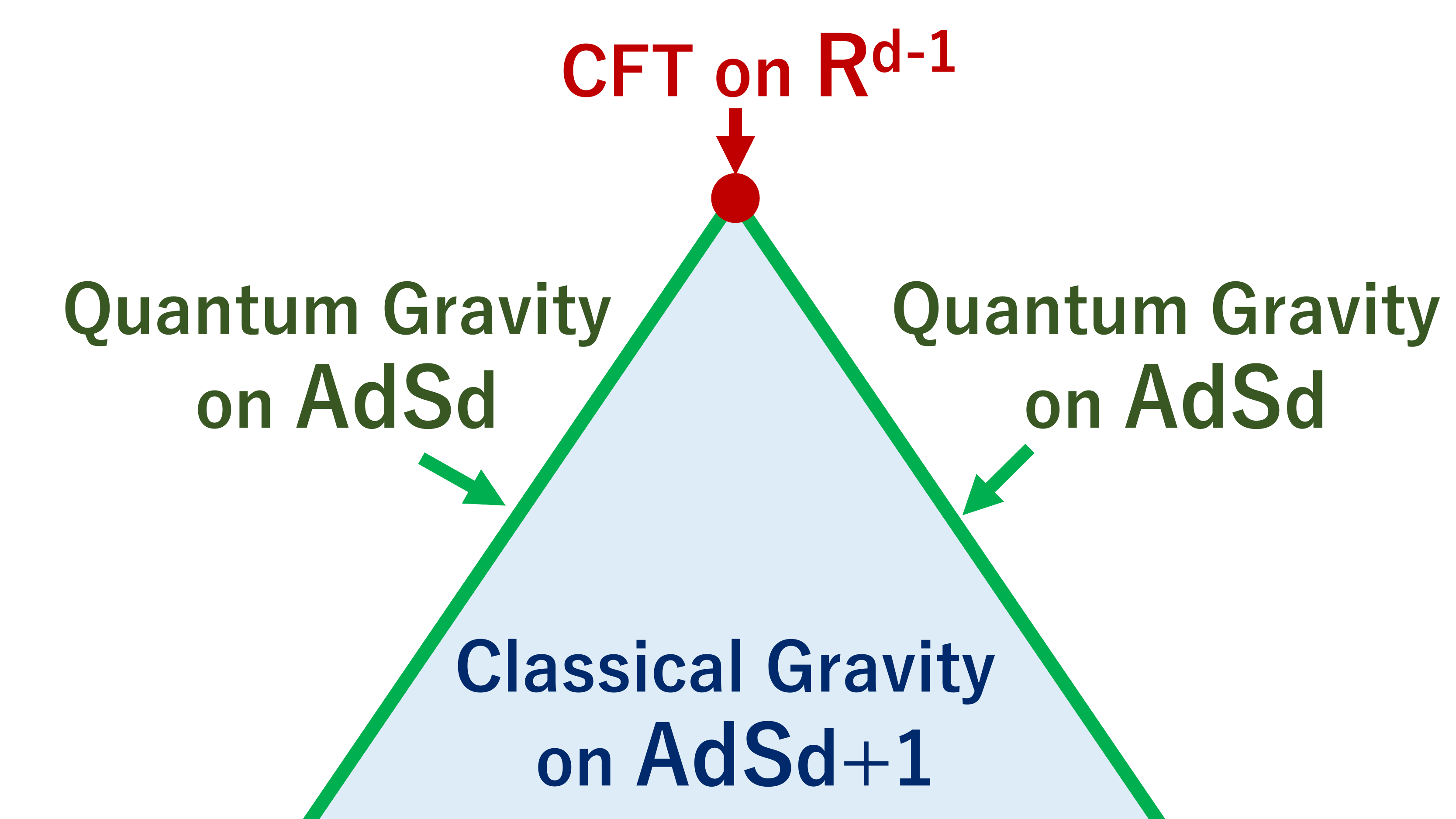}
  \caption{A sketch of wedge holography which argues that the gravity on a wedge region in AdS$_{d+1}$ is dual to a $d-1$ dimensional CFT on the codimension two spacetime given by the tip of the wedge.} 
\label{fig:wedgeh}
\end{figure}

The codimension two nature of the celestial holography looks mysterious for those who are familiar with normal holographic dualities such as the AdS/CFT. Recently, as a generalization of AdS/CFT, a new type of codimension two holography, called wedge holography, has been found in \cite{Akal:2020wfl} and studied further in \cite{Miao:2020oey,Miao:2021ual}. As sketched in Fig.\ref{fig:wedgeh}, the wedge holography argues that the gravity on a $d+1$ dimensional wedge region in AdS$_{d+1}$ is dual to a $d-1$ dimensional CFT on the $d-1$ dimensional tip of the wedge. We impose the Neumann boundary condition on $d$ dimensional boundaries of the wedge, so called the end of the world-branes (EOW branes). We can understand this as a small width limit of the AdS/BCFT \cite{Takayanagi:2011zk,Fujita:2011fp,Karch:2000gx}. Alternatively, we can also 
 understand the wedge holography via a double holography in the light of brane-world holography \cite{Randall:1999ee,Randall:1999vf,Gubser:1999vj,Karch:2000ct,Almheiri:2019hni} as follows. The $d+1$ dimensional gravity on the wedge is dual to a quantum gravity on the two $d$ dimensional EOW branes via the brane-world holography, which is further dual to a $d-1$ dimensional CFT on the tip via the standard holography.

\begin{figure}[hhh]
  \centering
  \includegraphics[width=12cm]{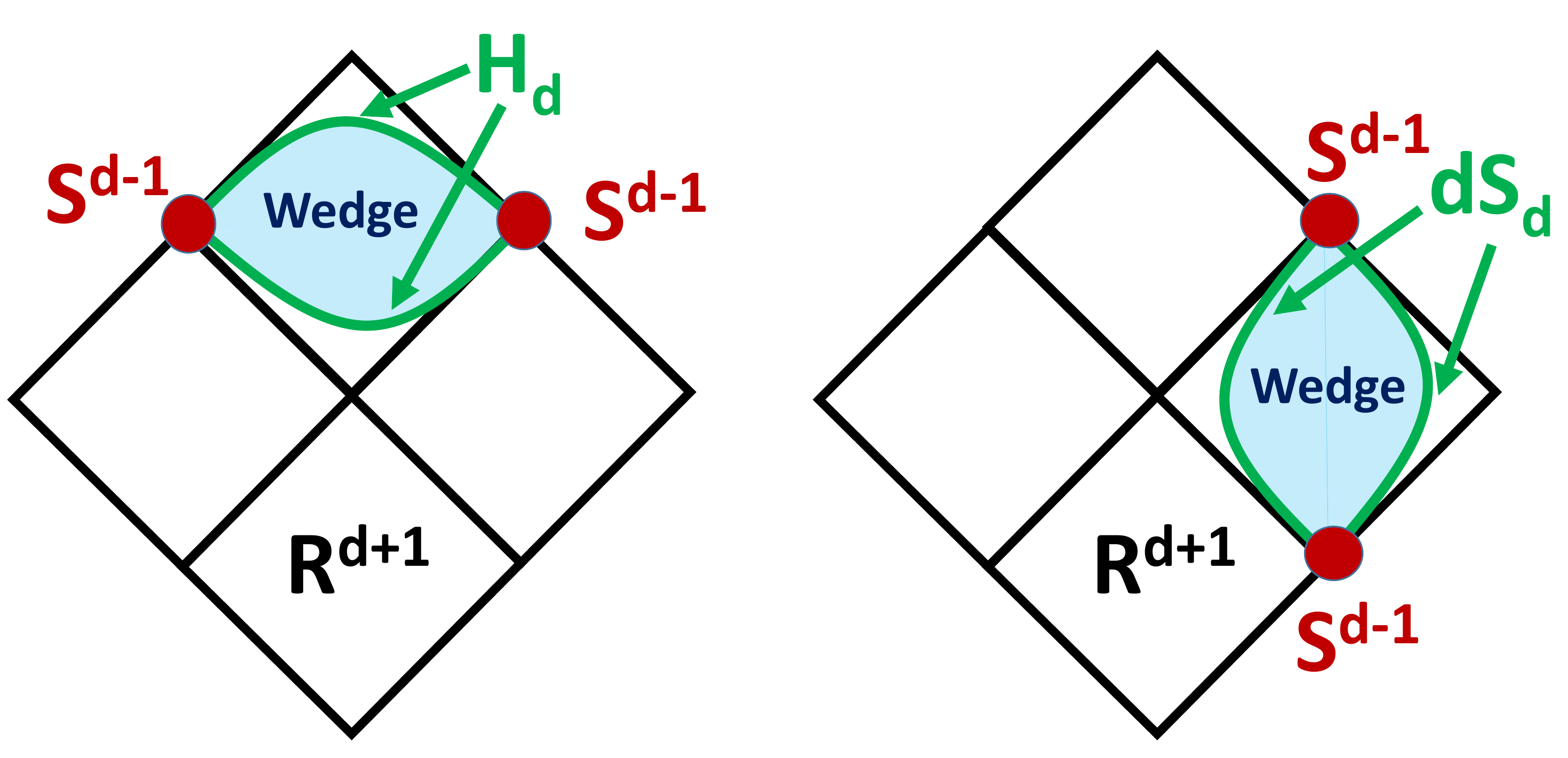}
  \caption{Sketches of two types of codimension two holographic dualities in flat space. The whole diamond describes a $d+1$ dimensional flat spacetime. The left and right panel describe the hyperbolic and de Sitter sliced wedges (blue regions) surrounded by two end of the world-brane (green surfaces), respectively. We argue that each of them is dual to a CFT on the $d-1$ dimensional sphere (red points).} 
\label{fig:wedgeah}
\end{figure}

Motivated by this, the main purpose of this paper is to explore if we can interpret the celestial holography as an extension of wedge holography to gravity on a flat spacetime. We consider two new classes of wedge holography depicted in 
Fig.\ref{fig:wedgeah}. One is a hyperbolic sliced wedge region and the other is a de Sitter sliced wedge region, both of which are surrounded by two space-like or time-like EOW branes, respectively. We argue that each of them is dual to a CFT on the $d-1$ dimensional sphere, situated at the tip of the wedge. The former might be interpreted as a product of lower dimensional AdS/CFT duality for Euclidean AdS geometries, though the product is now taken in the time direction as opposed to the standard wedge holography in  \cite{Akal:2020wfl}. The latter may be regarded as a product of lower dimensional dS/CFT, where the product is taken in the spacial direction\footnote{For an earlier study of a relation between celestial holography to the dS/CFT refer to \cite{Liu:2021tif}.}. We will examine these new holographic dualities by calculating the entanglement entropy, partition function and two point functions. Finally we will approach the celestial holography by combining these two dualities. 

This paper is organized as follows. In section two, we explain hyperbolic and de Sitter slices of Minkowski spacetime and solutions of a free scalar field  with a delta functional source on a sphere at null infinity. In section three, we propose a wedge holography in the hyperbolic patch and present evidences for this duality. In section four, we propose a wedge holography in the de Sitter patch and present evidences for this. In section five, we will try to interpret the celestial holography by combining the wedge holography in the hyperbolic slices and that in the de Sitter slices. In section six, we will summarize conclusions and discuss future problems. In appendix A, we briefly present useful identities related to Legendre functions. In appendix B, we describe minimal surfaces and geodesic length in hyperbolic spaces. In appendix C, we describe extreme surfaces and geodesic length in de Sitter spaces. In appendix D, we present detailed calculations of scalar modes in the de Sitter sliced wedges.

\section{Hyperbolic and de Sitter Slices of Flat Spacetime}
\label{sec:setup}
We start from a $d+1$ dimensional flat spacetime R$^{1,d}$:
\be
ds^2=-dT^2+dR^2+R^2d\Omega_{d-1}^2.
\ee
This is decomposed into two patches: the slices of hyperbolic spaces H$_d$ and de Sitter spaces dS$_d$, which suggest holographic properties \cite{deBoer:2003vf} (see also \cite{Ball:2019atb,Cheung:2016iub,Nguyen:2022zgs,Donnay:2022hkf}).

The hyperbolic slice is obtained by introducing the new coordinates
\ba
T=\eta\cosh\rho, \  \ R=\eta\sinh\rho.
\ea
This leads to the metric 
\ba
&& ds^2=-d\eta^2+\eta^2(d\rho^2+\sinh^2\rho d\Omega_{d-1}^2),\ \ \  \mbox{[hyperbolic patch]}, 
\label{hypatch}
\ea

On the other hand, the de Sitter slice is introduced by 
\ba
T=r\sinh t,\ \ \ R=r\cosh t,
\ea
which gives the metric 
\ba
&& ds^2=dr^2+r^2(-dt^2+\cosh^2 t d\Omega_{d-1}^2). \ \ \ \  \mbox{[de Sitter patch]}, 
\label{dspatch}
\ea

\begin{figure}
  \centering
  \includegraphics[width=8cm]{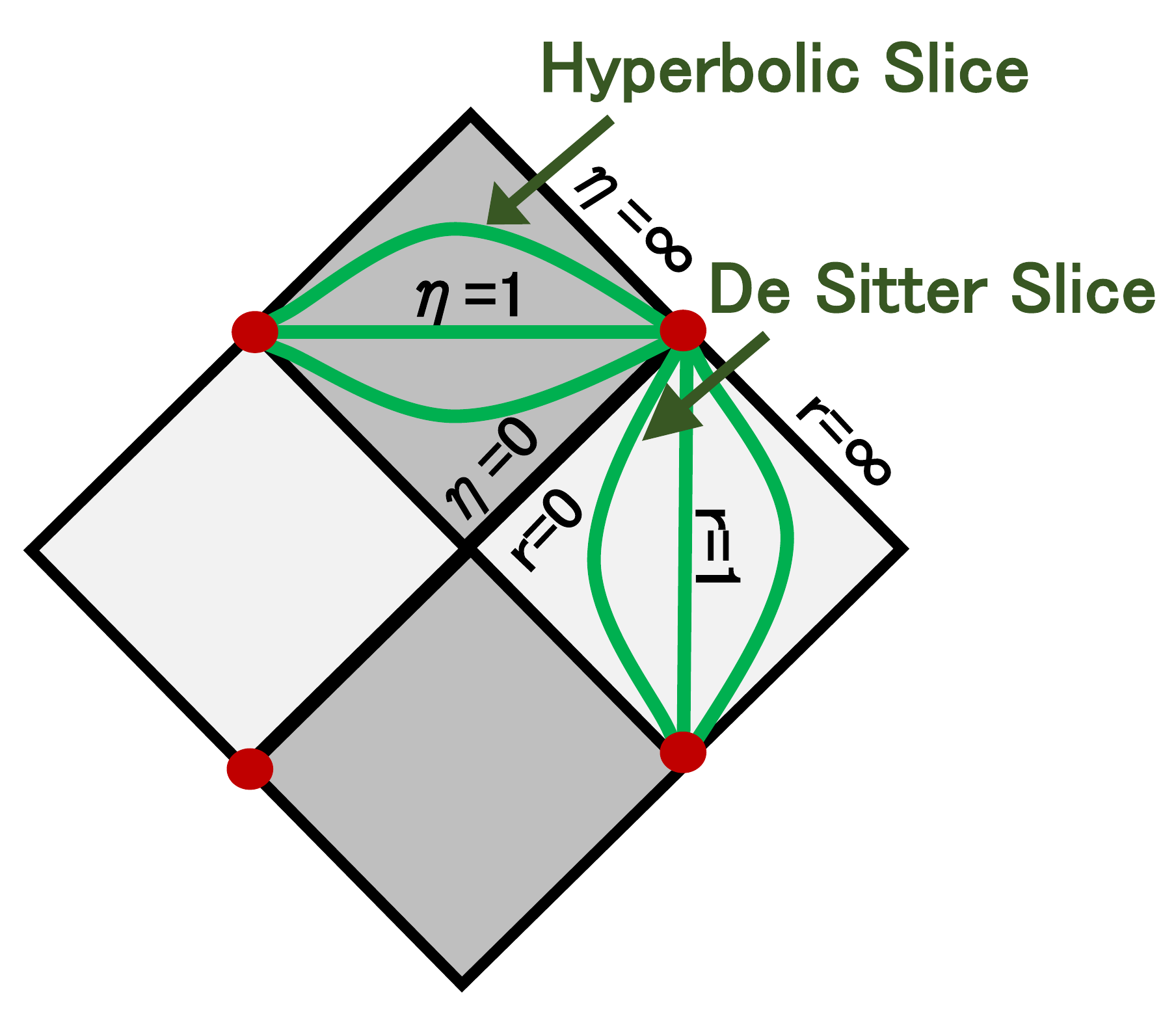}
  \includegraphics[width=6.5cm]{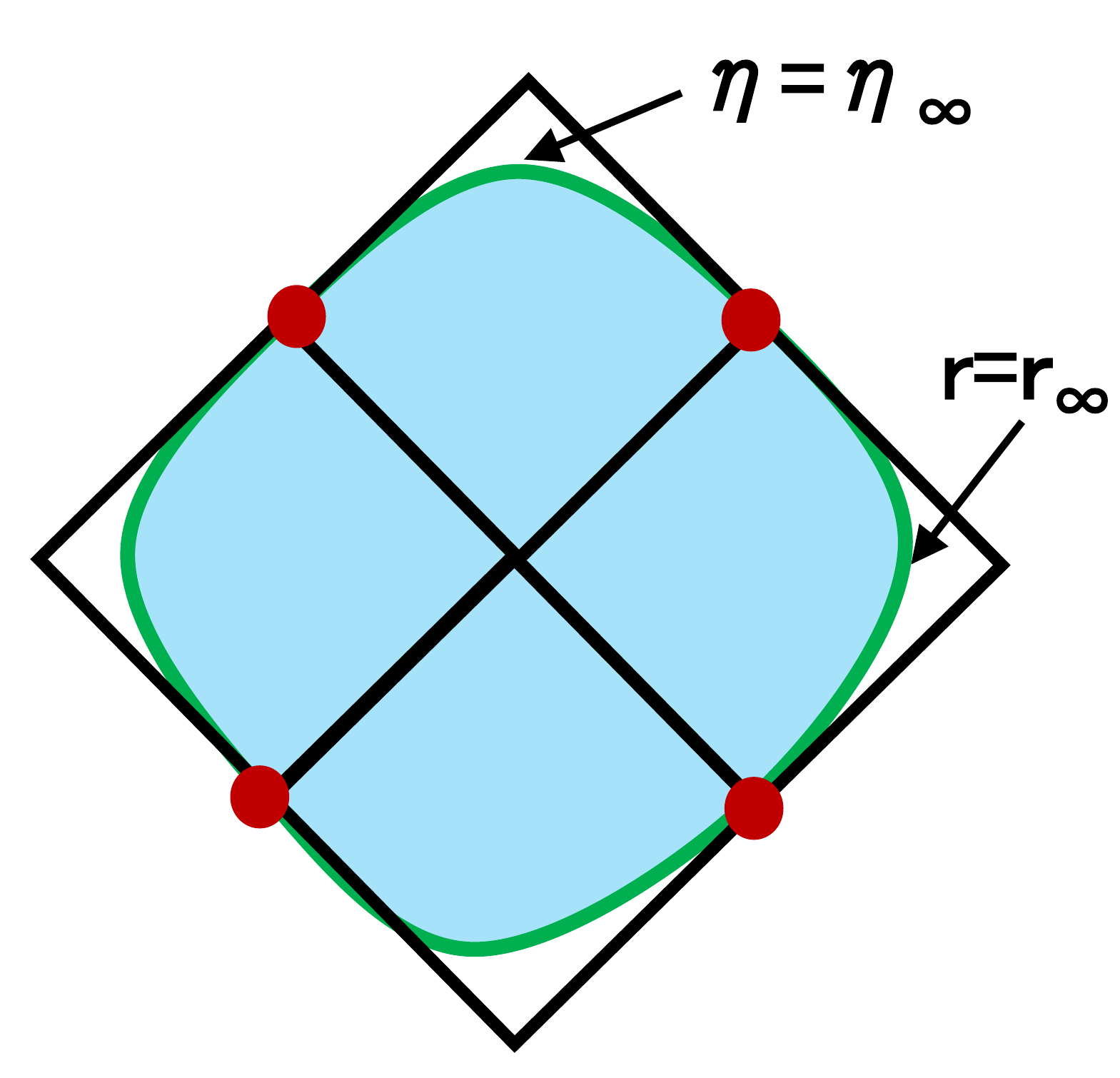}
  \caption{Hyperbolic and de Sitter slices in Minkowski Space (left) and its regularization (right).}
\label{fig:setupp}
\end{figure}

In these two patches, the radial coordinate $\eta$ and $r$ take the values $0\leq \eta<\infty$ and
$0\leq r<\infty$. By pasting the two patches along $\eta=0$ and $r=0$, we obtain the full four dimensional Minkowski spacetime as depicted in the left panel of Fig.\ref{fig:setupp}.

We introduce a regularization of the coordinates $\eta$ and $r$:
\ba
0\leq \eta\leq\eta_\infty,\ \ \ \ \ \ 0\leq r\leq r_\infty. \label{wedger}
\ea
This allows us to effectively reduce the hyperbolic patch and the de Sitter patch to H$_d$ and dS$_d$ via the compactification as analogous to the wedge holography for the AdS \cite{Akal:2020wfl}, which is a doubled version of the AdS/BCFT  \cite{Takayanagi:2011zk,Fujita:2011fp}.
If we extend the wedge holography to the $d+1$ dimensional Minkowski Space, one may be tempting to argue that a $d-1$ dimensional CFT on $S^{d-1}$ is dual to the gravity on the $d+1$ dimensional wedge region (\ref{wedger}).  As usual in the AdS/CFT \cite{Maldacena:1997re} and the dS/CFT \cite{Strominger:2001pn}, it is useful to introduce the UV cut off of the dual CFT, which is dual to the geometrical cut off 
\ba
\rho\leq\rho_\infty,\ \ \  t\leq t_\infty.  \label{cutoffb}
\ea
Below we will first study the hyperbolic and de Sitter slices separately by considering the wedge holography for each of them. After that we will discuss a connection between the celestial holography and the above wedge holography.

\subsection{Scalar field in hyperbolic patch}

Consider perturbations of a real scalar field $\Psi$ in the flat space, which are expected to be dual to scalar operator excitations in the dual CFT on the sphere in our wedge holography. We focus on the four dimensional gravity case i.e. $d+1=4$ just for simplicity. We write the two dimensional sphere metric as $d\Omega_2^2=d\theta^2+\sin^2\theta d\phi^2$.

We assume a massive free scalar field $\Psi$ given by the action 
\ba
I_{scalar}=\frac{1}{2}\int dx^4\s{-g}\left[-g^{\mu\nu}\de_\mu\Psi\de_\nu\Psi-m^2\Psi^2\right].
\ea
The equation of motion reads
\ba
\frac{1}{\s{-g}}\de_\mu\left(\s{-g}g^{\mu\nu}\de_\nu \Psi\right)-m^2\Psi=0.  \label{EOM}
\ea

 In the hyperbolic patch (\ref{hypatch}), the equation of motion of the scalar field (\ref{EOM}) is written as (see e.g. \cite{Raclariu:2021zjz})
 \ba
 -\de_\eta^2\Psi-\frac{3}{\eta}\de_\eta\Psi+\frac{1}{\eta^2}\left(\de_\rho^2\Psi+2\coth\rho\de_\rho\Psi\right)
 +\frac{1}{\eta^2\sinh^2\rho}\Delta_2\Psi-m^2\Psi=0,
 \ea
where $\Delta_2$ is the Laplacian on the two dimensional sphere. We can solve this by decomposing the solution as follows
\ba
\Psi(\eta,\rho,\theta,\phi)=f_p(\eta)g_{p,l}(\rho)Y_{lm}(\theta,\phi),
\ea
where the functions $f_p$, $g_{p,l}$ and $Y_{lm}$ satisfy
\ba
&& \left( -\de_\eta^2-\frac{3}{\eta}\de_\eta+\frac{p^2}{\eta^2}-m^2\right)f_p(\eta)=0,\no
&& \left(\de_\rho^2+2\coth\rho\de_\rho-\frac{l(l+1)}{\sinh^2\rho}-p^2\right)g_{p,l}=0,\no
&& \Delta_2 Y_{lm}=-l(l+1)Y_{lm}.
\ea
The first equation is explicitly solved as 
\ba
 f_p(\eta)=\ap\frac{I_{\s{1+p^2}}(m\eta)}{\eta}+\beta\frac{K_{\s{1+p^2}}(m\eta)}{\eta},\label{fpa}
\ea
 where $\ap$ and $\beta$ are arbitrary constants. 
 The solution to the second one reads
\ba
g_{p,l}(\rho)=\frac{1}{\sinh\rho} \cdot Q^{\s{1+p^2}}_l(\coth\rho),
\ea
 where $Q$ is the associated Legendre function. We chose Legendre $Q$ function instead of Legendre $P$ function because we require a smooth behavior at 
 $\rho=0$. Finally the function $Y_{lm}$ is the standard spherical harmonics
 (\ref{spharmo}).

\subsection{Scalar field in de Sitter patch} 
 
To obtain the solutions in the de Sitter patch (\ref{dspatch}),  we have only to replace the coordinate as
\ba
\eta=-ir,\ \ \ \rho=t-\frac{\pi}{2}i.  \label{cods}
\ea
This leads to the solution 
\ba
\Psi(r,t,\theta,\phi)=\ti{f}_p(r)\ti{g}_{p,l}(t)Y_{lm}(\theta,\phi),\label{dswave}
\ea
where each function reads
\ba
&& \ti{f}_p(r)=-\ap e^{\pi i\s{1+p^2}}\cdot\frac{J_{\s{1+p^2}}(mr)}{r}-\beta\frac{\pi i}{2} e^{-\pi i\s{1+p^2}}\cdot\frac{H^{(1)}_{\s{1+p^2}}(mr)}{r},\label{frti}\\
&& \ti{g}_{p,l}(t)=\frac{1}{\cosh t}\cdot Q^{\s{1+p^2}}_l(\tanh t).
\ea
Note that if we goes from $t=\infty$  to $t=-\infty$, the function $\ti{g}_{p,l}$
 gives the factor $(-1)^l$ because $Q^\mu_\nu(z)$ is given by $z^{-\mu-\nu-1}$ times an even function of $z$. This explains that  the future celestial sphere is related to the past one via the anti-podal map $\theta\to \theta+\pi$.

\subsection{Solution with a delta-functional source on the sphere}

We input an delta functional source of the scalar field at $(\theta_0,\phi_0)$ on S$^2$. In the hyperbolic slice, this corresponds to the following scalar field perturbation:
\ba
&& \Psi^{h}_0(\eta,\rho,\theta,\phi)\no
&& =f_p(\eta)\cdot \frac{{\cal N}}{\sinh\rho}\sum_{l=0}^\infty\sum_{m=-l}^l Y^*_{lm}(\theta_0,\phi_0)
Q^{\s{1+p^2}}_l(\coth\rho)Y_{lm}(\theta,\vp)\no
&&=f_p(\eta)\cdot \frac{{\cal N}}{\sinh\rho}\sum_{l=0}^\infty \left(\frac{2l+1}{4\pi}\right)
P_{l}(\cos\gamma)Q^{\s{1+p^2}}_l(\coth\rho),\label{fullpsi} 
\ea
where we employed the additivity formula (\ref{additive}) in the final line and we defined $\gamma$ by 
\ba
\cos\gamma=\cos\theta_0\cos\theta+\sin\theta\sin\theta_0\cos(\vp-\vp_0).
\label{gammadef}
\ea
We will choose the normalization factor ${\cal N}$ as ${\cal N}=e^{-\pi i\s{1+p^2}}\frac{1}{\Gamma(\s{1+p^2})}.$

In the $\rho\to\infty$ limit, using (\ref{expone}) and (\ref{exptwo}), we find 
\ba
&& \Psi^{h}_0(\eta,\rho,\theta,\phi)\no
&& \to f_p(\eta)\cdot\left[e^{\left(\s{1+p^2}-1\right)\rho}
\delta^2(\Omega-\Omega_0)+e^{-\left(\s{1+p^2}+1\right)\rho}\frac{\s{1+p^2}}{4\pi}\left(\frac{1-\cos\gamma}{2}\right)^{\Delta}\right],\no \label{bdybea}
\ea
which indeed gives the delta-functionally localized source with the correct $\rho$ dependence 
$e^{(\Delta-d)\rho}=e^{\left(\s{1+p^2}-1\right)\rho}$ for a source term
in AdS$_3/$CFT$_2$ by identifying the dimension of dual scalar operator $\Delta$ as
\ba
\Delta=1+\s{1+p^2}.  \label{confghw}
\ea

Moreover, we can show that the full expression of (\ref{fullpsi}) is expressed in the form \cite{deBoer:2003vf}:
\ba
\Psi^{h}_0(\eta,\rho,\theta,\phi)=\frac{\Delta-1}{4\pi}\cdot f_p(\eta)\cdot  
\frac{1}{(\cosh\rho-\cos\gamma\sinh\rho)^\Delta}.\label{phiH}
\ea
Indeed, we can prove the following expansion:
\ba
\left(\cosh\rho-\cos\gamma\sinh\rho\right)^{-\Delta}=\sum_{l=0}^\infty c_lP_l(\cos\gamma),
\ea
where $c_l$ can be found from the integral formula  (\ref{intfora}) as follows:
\ba
c_l=\frac{e^{-\pi i(\Delta-1)}(2l+1)}{\Gamma(\Delta)}\cdot \frac{1}{\sinh\rho}\cdot Q^{\Delta-1}_{l}(\coth\rho).
\ea

We can analytically continue the above analysis to the de Sitter slices via the coordinate transformation(\ref{cods}).
\ba
&& \Psi^{ds}_0(r,t,\theta,\phi)=\ti{f}_p(r)\cdot \frac{{\cal N}}{\cosh t}\sum_{l=0}^\infty \left(\frac{2l+1}{4\pi}\right)
P_{l}(\cos\gamma)Q^{\s{1+p^2}}_l(\tanh\rho).\label{fullpsids} 
\ea
In the $t\to \infty$ limit,  we find
\ba
&& \Psi^{ds}_0(r,t,\theta,\phi)\no
&& \to i\ti{f}_p(r)\cdot\left[e^{\left(\s{1+p^2}-1\right)\left(t-\frac{\pi}{2}i\right)}
\delta^2(\Omega-\Omega_0)+e^{-\left(\s{1+p^2}+1\right)\left(t-\frac{\pi}{2} i\right)}\frac{\s{1+p^2}}{4\pi}\left(\frac{1-\cos\gamma}{2}\right)^{\Delta}\right].\no \label{bdybeads}
\ea
The full function can be written as 
\ba
\Psi^{ds}_0(r,t,\theta,\phi)=\frac{\Delta-1}{4\pi}e^{\frac{\pi}{2}i(1-\Delta)}\cdot \ti{f}_p(r)\cdot  
\frac{1}{(\sinh t-\cos\gamma\cosh t+i\ti{\ep})^\Delta},\label{phidS}
\ea
where $\ti{\ep}$ is the regularization of $i\ep$ prescription \cite{deBoer:2003vf}.

\section{Wedge holography for hyperbolic slices}\label{sec:HP}

First we consider a wedge holography for hyperbolic slices depicted in the left panel of Fig.\ref{fig:wedgeah}. We specify the $d+1$ dimensional wedge $W^h$ by restricting the coordinate $\eta$ to the range 
\ba
\eta_1\leq \eta\leq \eta_2, \label{etawdge}
\ea
in the coordinate (\ref{hypatch}). We will impose the Neumann boundary condition 
on the two EOW branes $Q^{h(1)}$ and $Q^{h(2)}$ each at $\eta=\eta_1$ and $\eta=\eta_2$, given by 
\ba
K_{ab}-h_{ab}K=-Th_{ab},\label{NBY}
\ea
where $K_{ab}$ is the extrinsic curvature (we choose the normal vector $n^a$ is out-going) and $T$ is the tension of EOW brane. Indeed we can confirm that the boundary condition (\ref{NBY}) is satisfied by setting
the values of each tension to be 
\ba
T^{h(i)}=\frac{d-1}{d}K^{h(i)}=\frac{d-1}{\eta_{i}}, \label{NBYa}
\ea
where $i=1,2$ labels the two EOW branes.  

By extending the wedge holography in the AdS space \cite{Akal:2020wfl}, we argue that the $d+1$ dimensional gravity on the wedge $W^h$ (\ref{etawdge}) is dual to a $d-1$ dimensional CFT on the sphere S$^{d-1}$ at the tip $\rho\to \infty$.
We introduce the cut off $\rho=\rho_\infty$ as in (\ref{cutoffb}). Below we will give evidences for this new wedge holography by evaluating the partition function, holographic entanglement entropy and scalar field perturbation. Note that each hyperbolic slice H$_d$ at a fixed value of $\eta$ has the $SO(1,d)$ symmetry, which is the Lorentz symmetry in the original $d+1$ dimensional Minkowski spacetime. This symmetry matches with the conformal symmetry of the Euclidean CFT on S$^{d-1}$. In particular, at $d=3$, this is enhanced to a pair of Virasoro symmetries, which origins from the superrotation symmetry in 
R$^{1,3}$, being identified with the conformal symmetry of a dual two dimensional CFT.

Moreover, the results we will obtain below imply that the dual CFT on S$^{d-1}$ is non-unitary. This is not surprising because we added a time-like interval (\ref{etawdge}) as an internal direction, orthogonal to the hyperbolic space H$_d$, in spite that we can apply the standard AdS/CFT to each slice. Instead, this is analogous to the dS/CFT, where the dual CFT is expected to be non-unitary based on the analysis of central charge analysis \cite{Maldacena:2002vr} and explicitly known examples of the dS/CFT are non-unitary 
\cite{Anninos:2011ui,Cotler:2019nbi,Hikida:2021ese,Hikida:2022ltr}.

\subsection{Partition function}

The gravity action is written as follows:
\ba
I_G=\frac{1}{16\pi G_N}\int_{W^h} \s{-g}R+\frac{1}{8\pi G_N}\left[\int_{Q^{h(1)}}\s{\gamma}(K^{h(1)}-T^{h(1)})-
\int_{Q^{h(2)}}\s{\gamma}(K^{h(2)}-T^{h(2)})\right].\no
\ea
To evaluate the on-shell action,  we note the vanishing curvature $R=0$ and 
\ba
&&\int_{Q^{h(i)}}\s{\gamma}=\eta^d_{i}\omega_{d-1}\int^{\rho_\infty}_0 d\rho~ \sinh^{d-1} \rho,\ \  \ (i=1,2),
\ea
where we defined
\be
\omega_{d-1}=\frac{d\pi^{\frac{d}{2}}}{\Gamma(\frac{d}{2}+1)},\label{omegavol}
\ee
which is the volume of a unit sphere in d-1 dimension. 

By setting,
\be
J_d=\int^{\rho_\infty}_0 d\rho~ \sinh^{d} \rho,
\ee
and plugging (\ref{NBYa}), we obtain on-shell action as follows:
\be
I_G=-\frac{1}{8\pi G_N}(\eta^{d-1}_2-\eta^{d-1}_1)\omega_{d-1}J_{d-1}.
\label{Haction}
\ee
Note that $J_d$ obeys the recursion relation
\ba
J_d&=&\frac{1}{d}\sinh^{d-1}\rho_{\infty}\cosh \rho_{\infty} -\frac{d-1}{d}J_{d-2}.
\ea
Below we will explicitly evaluate the on-shell action for $d=3,4,5$.

\subsubsection{$d=3$ Case}

When $d=3$ we explicitly obtain
\ba
I^{d=3}_G=\frac{\eta^{2}_2-\eta^2_1}{16G_N}\left(-e^{2\rho_\infty}+4\rho_\infty\right).
\ea
By regarding the geometrical cut off $\rho_\infty$  in H$_d$ as the UV cut off $\ep$ in the dual two dimensional CFT on S$^2$ by identifying
\ba
\ep = e^{-\rho_\infty}, \label{cutoffrho}
\ea
we obtain
\ba
I_G= -\frac{\eta^2_2-\eta^2_1}{16G_N\ep^2}
-\frac{\eta^2_2-\eta^2_1}{4G_N}\log\ep.
\ea

This can be comparable to the standard CFT result that the sphere partition function of two dimensional CFT with the central charge $c$ reads \cite{Duff:1993wm,Henningson:1998gx}
\ba
Z_{CFT}\sim e^{\frac{A}{\ep^2}-\frac{c}{3}\log\ep},\label{partwocon}
\ea
where $A$ is non-universal constant, while the $\log\ep$ term is universal as this is fixed by the conformal anomaly. By equating this as $Z_{CFT}=e^{iI_G}$, we can estimate the central charge:
\ba
c=i\frac{3(\eta^2_2-\eta^2_1)}{4G_N}.  \label{centhyp}
\ea

\subsubsection{$d=4$ Case}

For $d=4$, the on-shell action reads
\be
I_G=\frac{\eta^3_2-\eta^3_1}{G_N}\left[-\frac{1}{48}e^{3\rho_{\infty}}+\frac{3}{16}e^{\rho_{\infty}} \right]
\ee
Using (\ref{cutoffrho}), we obtain
\be
I_G=-\frac{\eta^3_2-\eta^3_1}{48G_N\ep^3}
+\frac{3(\eta^3_2-\eta^3_1)}{16G_N\ep}.
\ee
This is expected to be dual to a three dimensional CFT.
The absence of logarithmic term in the gravity on-shell action is consistent with the well-known fact that there is no conformal anomaly in odd dimensional CFTs.

\subsubsection{$d=5$ Case}

For $d=5$, we obtain
\be
I_G=\frac{\pi(\eta^4_2-\eta^4_1)}{16G_N}\left[-\frac{1}{12}e^{4\rho_\infty}+\frac{2}{3}e^{2\rho_\infty}-2\rho_{\infty}\right].
\ee
In the same way as before, we can rewrite this as follows:
\be
I_G= \frac{\pi(\eta^4_2-\eta^4_1)}{2G_N}\left[-\frac{1}{96\ep^4}+\frac{1}{12\ep^2}+\frac{1}{4}\log\ep\right].
\ee

Now we would like to compare this result to the CFT one. The 4-sphere partition function with central charges $a,c$ satisfies the following equation \cite{Cardy:1988cwa,Duff:1993wm,Henningson:1998gx};
\ba
\ep\frac{d}{d\ep}\log Z_{CFT} 
&=&-\frac{1}{2\pi}\ev{\int d^4x\sqrt{g} T^{\mu}_{\mu}}_{S^4}\no
&=&-\frac{1}{2\pi}\int d^4x\sqrt{g}\left(\frac{a}{8\pi}\tilde{R}_{\mu\nu\rho\sigma}R^{\mu\nu\rho\sigma}-\frac{c}{8\pi}W_{\mu\nu\rho\sigma}W^{\mu\nu\rho\sigma} \right)\no
&=&-2a\chi(S^4)\no
&=&-4a.\label{4dpart}
\ea
In the third line, we use
\be
\ev{T^{\mu}_{\mu}}=\frac{a}{8\pi}\tilde{R}_{\mu\nu\rho\sigma}R^{\mu\nu\rho\sigma}-\frac{c}{8\pi}W_{\mu\nu\rho\sigma}W^{\mu\nu\rho\sigma}.
\ee
And in the forth line, we use the Euler characteristic class in four dimensional manifold
\be
\frac{1}{32\pi^2}\int d^4x\sqrt{g}\tilde{R}_{\mu\nu\rho\sigma}R^{\mu\nu\rho\sigma}=\chi(M),
\ee
and the fact that the Weyl tensor $W$ vanishes in $S^4$. By solving (\ref{4dpart}), we obtain the logarithmic part of $Z_{CFT}$, 
\be
\log Z_{CFT}=4a\log\ep+\mathrm{(other~parts)}.\label{actionanofour}
\ee

By equating this as $Z_{CFT}=e^{iI_G}$, we can estimate the central charge:
\be
a=i\frac{\pi(\eta^4_2-\eta^4_1)}{32G_N}.  \label{centfour}
\ee

\subsection{Holographic entanglement entropy}

Now we would like to calculate the holographic entanglement entropy \cite{Ryu:2006bv,Ryu:2006ef,Hubeny:2007xt}, which is given by 
\ba
S_A=\frac{\mbox{Area}(\Gamma_A)}{4G_N},  \label{HEE}
\ea
where $\Gamma_A$ is the extremal surface which ends on the boundary of $A$
i.e. $\de\Gamma_A=\de A$.

The $d-1$ dimensional extremal surface\footnote{Refer to \cite{Kapec:2016aqd} for earlier calculations of entanglement entropy in asymptotically flat spacetime, where the entangling surface lies at null infinity. On the other hand, in our case, we are considering a different quantity, namely the entanglement entropy of two dimensional CFT on a celestial sphere, where the dual extremal surface extends from the null infinity to the bulk of flat space. } which computes the holographic entanglement entropy in our $d+1$ dimensional wedge $W^h$ is given by a family of the minimal area surfaces in the hyperbolic spaces H$_d$ parameterized by the time coordinate $\eta$ in the range (\ref{etawdge}). Such an extreme surface is time-like and its area takes a pure imaginary value, as is common to the holographic entanglement entropy in dS/CFT correspondence \cite{Narayan:2015vda, Sato:2015tta,Hikida:2022ltr}. 

At a fixed value of $\eta$, this is given by the standard minimal surface $\gamma^H_A$ in H$_d$ calculated in Appendix \ref{appendix:hdmin}. We take the metric of $H_d$ to be 
\ba
ds^2=d\rho^2+\sinh^2\rho (d\theta_1^2+\sin^2\theta_1d\Omega_{d-2}).  \label{hypbl}
\ea
The minimal area which stretches between $\theta=-\theta_0$ to $\theta=\theta_0$ is given by
\ba
A(\gamma^H_A)=\omega_{d-3}\cdot \int^{1}_{\frac{\delta}{L}}dy\frac{(1-y^2)^{\frac{d-4}{2}}}{y^{d-2}},
\ea
where the infinitesimally small cut off  $\delta$ and the minimal surface parameter $L$, are related to the cut off $\rho_\infty$ and $\theta_0$ via (\ref{CUTa}) and  (\ref{CUTb}).
We again denote the volume of unit $d$ dimensional sphere by $\omega_{d}$ as in (\ref{omegavol}).
For $d=3,4$ and $5$ we obtain the following results:
\ba
&& A(\gamma^H_A)_{d=3}=2\log\frac{2L}{\delta}=2\log (e^{\rho_{\infty}}\sin\theta_0),\no
&&  A(\gamma^H_A)_{d=4}=2\pi\left(\frac{L}{\delta}-1\right)=\pi\sin\theta_0 e^{\rho_\infty}-2\pi,\no
&& A(\gamma^H_A)_{d=5}=4\pi\left[\frac{L^2}{2\delta^2}-\frac{1}{2}\log\frac{2L}{\delta}-\frac{1}{4}\right]\no
&& =\frac{\pi}{2}\sin^2\theta_0 e^{2\rho_\infty}-2\pi\log\left(\frac{\sin\theta_0}{4}e^{\rho_\infty}\right)
-\pi\cos\theta_0(2-\cos\theta_0).
\ea

Thus the area of the full extremal surface in $W^h$ is given by 
\ba
A(\Gamma^H_A)=i\int^{\eta_2}_{\eta_1}\eta^{d-1}d\eta \cdot A(\gamma^{H}_A)=\frac{i}{d-1}(\eta^{d-1}_2-\eta^{d-1}_2)
 A(\gamma^{H}_A).\label{minareaH}
\ea

In this way the total expression of the holographic entanglement entropy reads
\ba
S_A=\frac{A(\Gamma^{H}_A)}{4G_N}=\frac{i(\eta^{d-1}_2-\eta^{d-1}_1)}{4(d-1)G_N}
 A(\gamma^{H}_A).
\ea

For $d=3$ we obtain
\ba
S_A=\frac{i(\eta_2^2-\eta_1^2)}{8G_N}\log\left(\frac{\sin^2\theta_0}{\ep^2}\right),
\ea
where we employed the relation between the CFT cut off $\ep$ and the gravity cut off 
$\rho_\infty$ (\ref{cutoffrho}). We can compare this with the standard result computed in a two dimensional CFT on S$^2$ with a central charge $c$, where the subsystem $A$ is taken to be an interval; $\theta^{(1)}\leq \theta_1\leq \theta^{(2)}$, given by \cite{Holzhey:1994we,Calabrese:2004eu}:
\ba
S_A=\frac{c}{6}\log\left[\frac{\sin^2\left(\frac{\theta^{(1)}-\theta^{(2)}}{2}\right)}
{\ep^2}\right].\label{CCfor}
\ea
This comparison tells us that the central charge takes the following imaginary value 
\ba
c=\frac{3i}{4G_N}(\eta^2_2-\eta^2_1).  \label{centrc}
\ea
This agrees with the value (\ref{centhyp}) computed from the partition function.

For $d=5$ we find by setting $\rho_\infty=-\log\ep$ (\ref{cutoffrho}):
\ba
S_A=i\frac{\pi(\eta^4_2-\eta^4_1)}{16G_N}\left[\frac{\sin\theta_0^2}{2\ep^2}
-2\log\left(\frac{\sin\theta_0}{4\ep}\right)-\cos\theta_0(2-\cos\theta_0)\right].
\ea
By comparing this with the standard formula in four dimensional CFTs \cite{Ryu:2006ef,Hung:2011xb}
\ba
S_A=c_0\ep^{-2}+4a\log\ep+O(1),  \label{EEano}
\ea
we can read off the value of the central charge $a$:
\ba
a=\frac{\pi i}{32G_N}(\eta^4_2-\eta^4_1).  \label{centrala}
\ea
Indeed this agrees with our previous evaluation from the partition function (\ref{centfour}).

\subsection{Scalar Field Perturbation and Two Point Functions}\label{sec:Hscalar}

Consider a free massive scalar field in the wedge geometry $W^h$, defined by $\eta_1\leq \eta\leq \eta_2$. We expect this is dual to scalar operators in the dual CFT on S$^{d-1}$.
We impose either the Dirichlet or Neumann boundary condition at the boundary $\eta=\eta_{1,2}$. As we will show below, there are infinitely many scalar modes dual to operators which have conformal dimension $\Delta=1+i\lambda_k$ with $\lambda_k$ real valued. This complex valued conformal dimension again suggests the non-unitary nature of the dual CFT as similar to the celestial holography \cite{Pasterski:2016qvg}.

\subsubsection{Dirichlet boundary condition}
\begin{figure}[hhh]
  \centering
  \includegraphics[width=8cm]{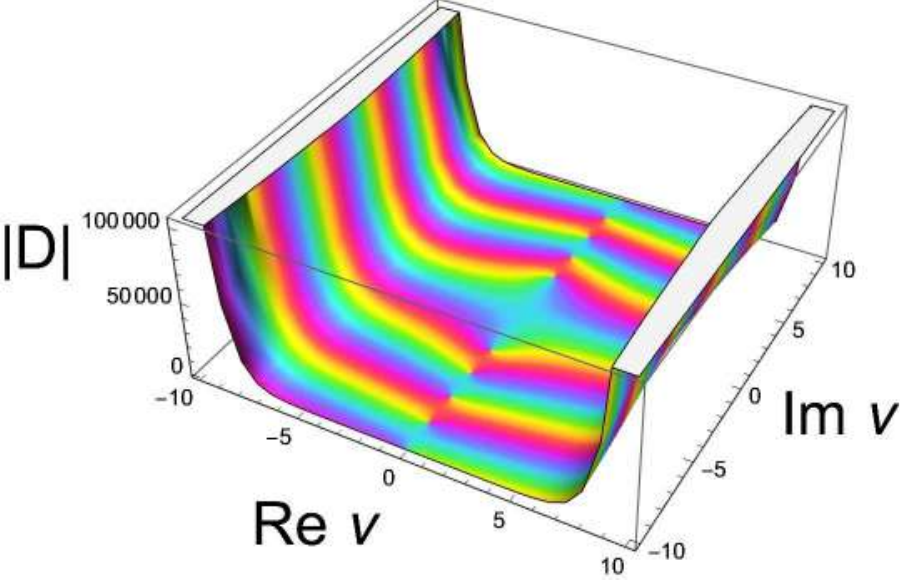}
   \includegraphics[width=6cm]{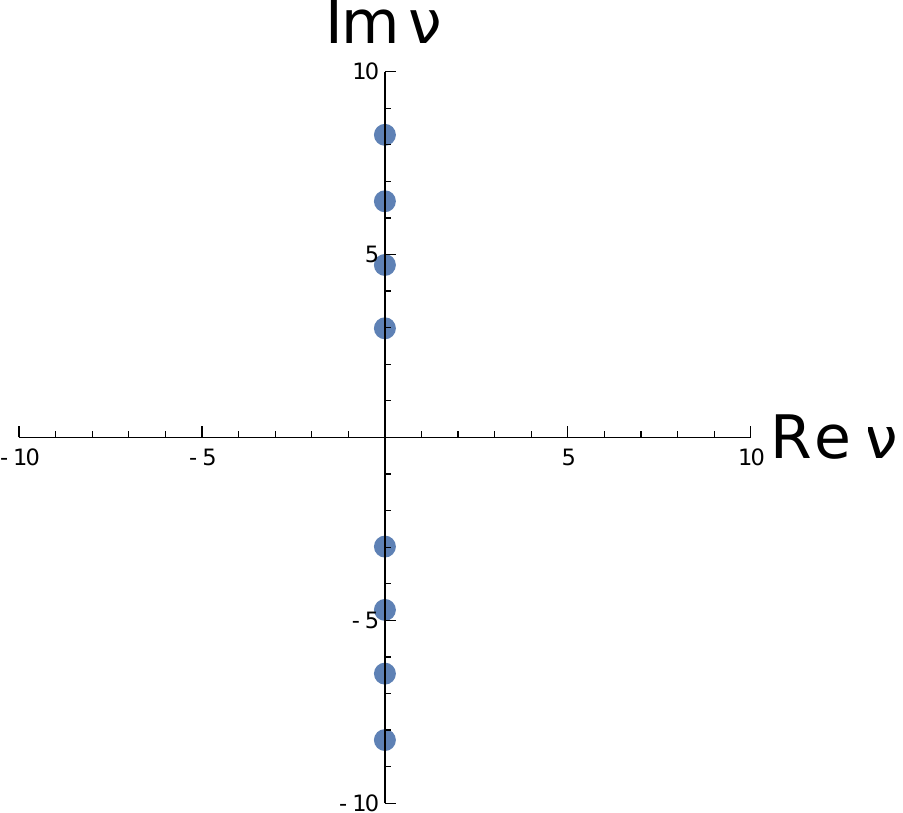}
  \caption{Plots of $D^h(\nu,x_1,x_2)$ for $x_1=1$ and $x_2=5$ as a function of $\nu$ (left) and plots of zero point of $D^h$ (right). The color of left figure represents $D$'s phase.}
\label{fig:zeroha}
\end{figure}

 We impose the Dirichlet boundary condition on the two EOW-branes:
\begin{equation}
\label{bdycondhya}
\begin{aligned}
f_p(\eta_1)&=0, \ \ \ f_p(\eta_2)&=0,
\end{aligned}
\end{equation}
where the function $f_p(\eta)$ was defined in (\ref{fpa}). Solving this boundary condition is equivalent to the search of values of $\nu=\s{1+p^2}$ which satisfy 
\ba
D^h(\nu,x_1,x_2)=I_{\nu}(x_1)K_{\nu}(x_2)-I_{\nu}(x_2)K_{\nu}(x_1)=0,  \label{dheq}
\ea
where $x_{1,2}= m \eta_{1,2}$. Solutions exist only when $\nu$ is pure imaginary and there are infinitely many discrete solutions as depicted in Fig.\ref{fig:zeroha}. We write the values of $\nu$ which satisfy (\ref{dheq}) as $\nu=i\lambda_k$.
Note that if $\nu=i\lambda_k$ is a solution, then its complex conjugate $\nu=-i\lambda_k$ is also a solution.
This shows that a bulk scalar with mass $m$ is dual to infinitely many scalar operators which have  
complex and discrete values of conformal dimension:
\ba
\Delta=1+i\lambda_k. \label{codima}
\ea

To see this property analytically, we take two limits: $\eta_2\to \infty$ and $\eta_1\to 0$. The first boundary condition can be written as
\be
f_p(\eta_2)=\ap\frac{I_{\nu}(m\eta_2)}{\eta_2}+\beta\frac{K_{\nu}(m\eta_2)}{\eta_2}=0.
\ee
In the limit $\eta_2\to \infty$, the first term diverges. Thus we must set $\ap=0$.
In order to satisfy the second condition, 
we require $\lim_{\eta_1\rightarrow0} K_{\nu}(m\eta_1)=0$. Recalling
\begin{align}
\label{modifiedBessel0limit}
    K_\nu(m\eta)\simeq \frac{\pi}{2} \frac{1}{\sin \nu \pi} 
    \left\{\frac{\left(\frac{m\eta}{2}\right)^{-\nu}}{\Gamma(1-\nu)}-\frac{\left(\frac{m\eta}{2}\right)^\nu}{\Gamma(1+\nu)}
    \right\}
    ~~~~(0<\eta\ll1),
\end{align}
it is obvious that $K_\nu(0)$ diverges if $\nu$ has a real part. Thus we set $\nu=i\lambda$ $(\lambda\in\mathbb{R})$.
\begin{align}
    K_{i\lambda}(m\eta)\simeq~
    \frac{\pi}{2} 
    \frac{
    \frac{e^{-i\lambda\log\frac{m\eta}{2}}}{\Gamma(1-i\lambda)}-\frac{e^{+i\lambda\log\frac{m\eta}{2}}}{\Gamma(1+i\lambda)}
    }
    {i\sinh \lambda \pi}
    =~\frac{\pi}{\sinh \lambda \pi}\mathrm{Im}\left(\frac{e^{-i\lambda\log\frac{m\eta}{2}}}{\Gamma(1-i\lambda)}\right).
\end{align}
In the last line we can see that 
infinitely many (but discrete) values of $\lambda$ satisfies the necessary condition at nonzero $\eta$. We can also see that the satisfactory values of $\lambda$ become continuous under $\eta\to0$ because $\log\frac{m\eta}{2}\to-\infty$.

\subsubsection{Neumann boundary condition}
\begin{figure}[hhh]
  \centering
  \includegraphics[width=8cm]{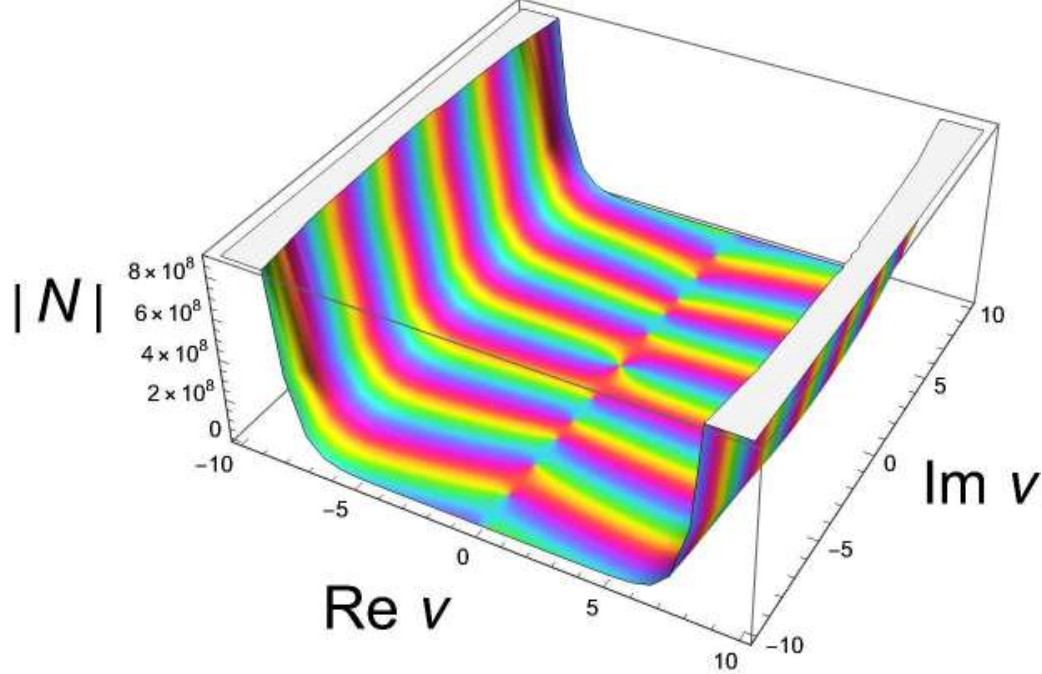}
   \includegraphics[width=6cm]{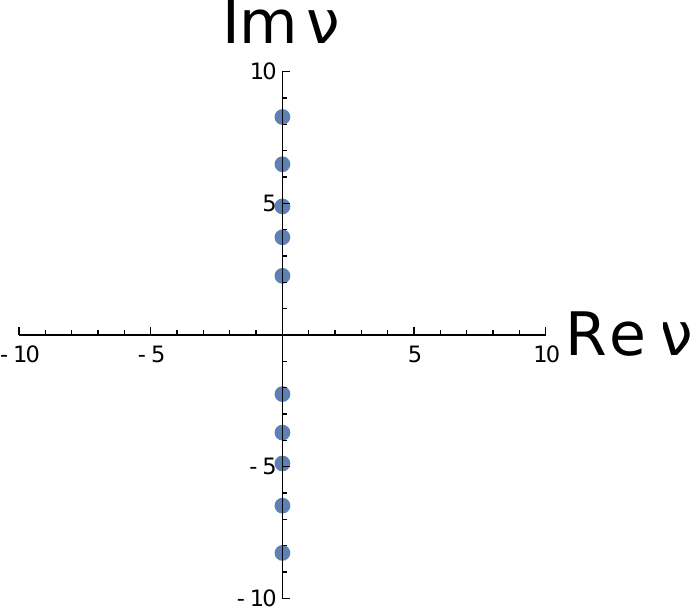}
  \caption{Plots of $N^h(\nu,x_1,x_2)$ for $x_1=1$ and $x_2=5$ as a function of $\nu$ (left) and plots of zero point of $N^h$ (right). The color of left figure represents $N$'s phase.}
\label{fig:h_neumann}
\end{figure}

Now we impose the Neumann boundary condition on the two EOW-branes:
\begin{equation}
\label{bcn}
\begin{aligned}
\partial_{\eta}f_p(\eta_1)&=0, \ \ \ \ 
\partial_{\eta}f_p(\eta_2)&=0,
\end{aligned}
\end{equation}
where the function $f_p(\eta)$ was defined in (\ref{fpa}).
By using the recurrence formula of modified Bessel function
\ba
\partial_x I_\nu(x)=\frac{1}{2}\left(I_{\nu+1}(x)+I_{\nu-1}(x)\right),\ \ \ 
\partial_x K_\nu(x)=-\frac{1}{2}\left(K_{\nu+1}(x)+K_{\nu-1}(x)\right),\no
\frac{I_\nu(x)}{x}=-\frac{1}{2\nu}\left(I_{\nu+1}(x)-I_{\nu-1}(x)\right), \ \ \
\frac{K_\nu(x)}{x}=\frac{1}{2\nu}\left(K_{\nu+1}(x)-K_{\nu-1}(x)\right),
\ea
we can write (\ref{bcn}) as follows:
\begin{align}
\ap\left( 
\frac{\nu+1}{2\nu}I_{\nu+1}(m\eta_a)\right.&+\left.\frac{\nu-1}{2\nu}I_{\nu-1}(m\eta_a)\right)
-\beta\left(
\frac{\nu+1}{2\nu}K_{\nu+1}(m\eta_a)+\frac{\nu-1}{2\nu}K_{\nu-1}(m\eta_a)
\right)=0.\nonumber
\end{align}
where $a=1,2$. This is equivalent to the search of values of $\nu=\s{1+p^2}$ which satisfy 
\ba
N^{h}(\nu,x_1,x_2)&=&\left\{(\nu+1)I_{\nu+1}(x_1)+(\nu-1)I_{\nu-1}(x_1)\right\}\left\{(\nu+1)K_{\nu+1}(x_2)+(\nu-1)K_{\nu-1}(x_2)\right\} \no
&&-\left\{(\nu+1)I_{\nu+1}(x_2)+(\nu-1)I_{\nu-1}(x_2)\right\}\left\{(\nu+1)K_{\nu+1}(x_1)+(\nu-1)K_{\nu-1}(x_1)\right\} \no
&=&0.
\label{EEE}
\ea
where $x_{1,2}= m \eta_{1,2}$. Solutions exist only when $\nu$ is pure imaginary and there are infinitely many discrete solutions as depicted in Fig.\ref{fig:h_neumann}. We write the values of $\nu$ which satisfy (\ref{EEE}) as $\nu=i\lambda_k$. Each mode is dual to a scalar operator with the conformal dimension (\ref{codima}).
Again, if $\nu=i\lambda_k$ is a solution, then its complex conjugate $\nu=-i\lambda_k$ is also a solution.

We take the limit $\eta_2\to \infty$ and $\eta_1\to 0$. The first equation can be written as
\begin{align}
\partial_{\eta}f_p(\eta_2)=&~m\ap\left( 
\frac{\nu+1}{2\nu}\frac{ I_{\nu+1}(m\eta_2)}{\eta_2}+\frac{\nu-1}{2\nu}\frac{ I_{\nu-1}(m\eta_2)}{\eta_2}\right) \no
&-m\beta\left(
\frac{\nu+1}{2\nu}\frac{ K_{\nu+1}(m\eta_2)}{\eta_2}+\frac{\nu-1}{2\nu}\frac{ K_{\nu-1}(m\eta_2)}{\eta_2}
\right)=0.
\end{align}
In the limit $\eta_2\to \infty$, the first term diverges. Thus we must take $\ap=0$.
The second equation can be written as
\be
\partial_{\eta}f_p(\eta_1)=m\beta\left(
\frac{\nu+1}{2\nu}\frac{K_{\nu+1}(m\eta_1)}{\eta_1}+\frac{\nu-1}{2\nu}\frac{K_{\nu-1}(m\eta_1)}{\eta_1}
\right)=0.
\ee
In order to satisfy this equation in the limit $\eta_1\rightarrow0$, it is necessary to impose
\be
\lim_{\eta_1\rightarrow0} \left\{(\nu+1)\frac{K_{\nu+1}(m\eta_1)}{\eta_1}+(\nu-1)\frac{K_{\nu-1}(m\eta_1)}{\eta_1}
\right\}=0.
\ee
In $0<\eta_1 \ll1$, applying the asymptotic form (\ref{modifiedBessel0limit}),
\begin{align}
    (\mathrm{inside~the ~limit})=& 
    (\nu+1)\frac{K_{\nu+1}(m\eta_1)}{\eta_1}+(\nu-1)\frac{K_{\nu-1}(m\eta_1)}{\eta_1} \no 
\simeq&-\frac{\pi}{2} \frac{1}{\sin \nu \pi}\frac{m}{2}
\left\{
\frac{(1+\nu)\left(\frac{m \eta_{1}}{2}\right)^{-2-\nu}}{\Gamma(-\nu)}
+\frac{(1-\nu)\left(\frac{m\eta_{1}}{2}\right)^{-2+\nu}}{\Gamma(\nu)}
\right\}.
\end{align}
It is obvious that the last line diverges if $\nu$ has real part. Thus we set $\nu=i\lambda$ $(\lambda\in\mathbb{R})$,
\begin{align}
    (\mathrm{inside~the~limit})
    \simeq 
    -\frac{\pi}{\sin{i\lambda \pi}}\frac{2}{m\eta_1^2}\mathrm{Re}\left[ (1-i\lambda) \frac{e^{i\lambda\log\frac{m\eta_1}{2}}}{\Gamma(i\lambda)}\right].
\end{align}
We can see that 
infinitely many (but discrete) values of $\lambda$ satisfies the necessary condition at nonzero $\eta_1$. We can also see that the satisfactory values of $\lambda$ become continuous under $\eta_1\to 0$ because $\log\frac{m\eta_1}{2}\to-\infty$.

\subsubsection{Two point function}
Now let us calculate the two point functions by extending the standard bulk to boundary relation in AdS/CFT \cite{Gubser:1998bc,Witten:1998qj} to our wedge holography. For this we evaluate the on-shell action of scalar field
\ba
I_{scalar}=i\int^{\eta_2}_{\eta_1}\frac{d\eta}{\eta}f_p(\eta)^2\cdot 
\sinh^2\rho_\infty\cdot \int d^2\Omega\Psi\de_\rho\Psi|_{\rho=\rho_\infty}.
\ea
Then it is obvious that we obtain the two point function of dual scalar operators at fixed value of $p$ where the product of the first and second term of 
(\ref{bdybea}) contributes:
\ba
\la O_p(\theta_1,\vp_1)O_p(\theta_2,\vp_2)\lb\propto \left(1-\cos\gamma_{12}\right)^{-\Delta},\label{correlationf}
\ea
where $\Delta$ is related to $p$ via (\ref{confghw}) and we also introduced $\gamma_{12}$ by
\ba
\cos\gamma_{12}=\cos\theta_1\cos\theta_2+\sin\theta_1\sin\theta_2\cos(\vp_1-\vp_2).
\label{gammadexf}
\ea
This agrees with the expected two point function of two dimensional CFT on S$^2$ by identifying $\Delta$ with the conformal dimension of the scalar operator $O_p$.

\section{Wedge holography for de Sitter slices} \label{sec:dS}

As the second wedge holography, we consider the $d+1$ dimensional wedge $W^{ds}$ by restricting the de Sitter sliced metric (\ref{dspatch}) to the region 
\ba
r_1\leq r\leq r_2,  \label{etawdgee},
\ea
as sketched in the right panel of Fig.\ref{fig:wedgeah}. The two boundaries $r=r_1$ and $r=r_2$ are the EOW branes $Q^{ds(1)}$ and $Q^{ds(2)}$, where we impose the Neumann boundary condition (\ref{NBY}). By solving this boundary condition, we obtain 
\ba
T^{ds(i)}=\frac{d-1}{d}K^{ds(i)}=\frac{d-1}{r_{i}}, \label{NBYbaa}
\ea
where $i=1,2$ labels the two EOW branes.  

We argue that the $d+1$ dimensional gravity on the wedge $W^{ds}$ (\ref{etawdgee}) is dual to a $d-1$ dimensional CFT on a $d-1$ dimensional sphere S$^{d-1}$. Even though there are two spheres situated at the tips of the wedge: $t=-\infty$ and $t=\infty$, we identify them via the antipodal mapping. We introduce the cut off $t=\pm t_\infty$ as in (\ref{cutoffb}). As in the hyperbolic case, each de Sitter slice dS$_d$ at a fixed value of $r$ has $SO(1,d)$ symmetry. This is the Lorentz symmetry in the original $d+1$ dimensional Minkowski spacetime and matches with the conformal symmetry of the dual Euclidean CFT on S$^{d-1}$. At $d=3$, this is again enhanced to a pair of Virasoro symmetries. This is the superrotation symmetry \cite{Barnich:2009se,Barnich:2010ojg} in  R$^{1,3}$ and is identified with the conformal symmetry of a dual two dimensional CFT.

Notice that this wedge holography can be regarded simply as a dS version of the wedge holography in the AdS \cite{Akal:2020wfl} because the wedge is defined by adding a spacial width to a dS$_d$. Therefore we again expect the dual CFT on S$^{d-1}$ is non-unitary being similar to the dS/CFT \cite{Strominger:2001pn,Maldacena:2002vr,Anninos:2011ui,Cotler:2019nbi,Hikida:2021ese,Hikida:2022ltr}. We will study the partition function, holographic entanglement entropy and scalar field perturbation to verify this wedge holography.

\subsection{Partition function}

The gravity action on our wedge region reads
\ba
I_G=\frac{1}{16\pi G_N}\int_{W^{ds}} \s{-g}R-\frac{1}{8\pi G_N}\left[\int_{Q^{ds(1)}}\s{-h}(K^{ds(1)}-T^{ds(1)})
-\int_{Q^{ds(2)}}\s{-h}(K^{ds(2)}-T^{ds(2)})\right].\no
\ea
We would like to limit the spacetime to be the half $0\leq t<\infty$. 

By noting
\ba
&&\int_{Q^{ds(i)}}\s{-h}=r^d_{i}\omega_{d-1}\int^{t_\infty}_0 dt~ \cosh^{d-1} t,
\ea
and introducing 
\be
I_{d}=\int^{t_\infty}_0 dt~ \cosh^{d} t,
\ee
we obtain on-shell action as follows:
\be
I_G=\frac{1}{8\pi G_N}(r_{2}^{d-1}-r_1^{d-1})\omega_{d-1}I_{d-1}.\label{dSaction}
\ee
We can find the recurrence formula as follows:
\ba
I_d&=&\frac{1}{d}\cosh^{d-1}t_{\infty}\sinh t_{\infty} +\frac{d-1}{d}I_{d-2}.
\ea
Below we would like to evaluate this explicitly for $d=3,4$ and $5$.

\subsubsection{$d=3$ Case}

For $d=3$, the on-shell action reads
\ba
I_G=\frac{r_2^2-r_1^2}{16G_N}\left(e^{2t_\infty}
+4  t_\infty\right).
\ea
By regarding the geometrical cut off $t_\infty$ in the dS$_3$ as the UV cut off $\ep$ in the dual two dimensional CFT on S$^2$ by identifying
\ba
\ep = e^{-t_\infty}, \label{cutoffto}
\ea
we obtain
\ba
I_G= \frac{r^2_2-r^2_1}{16G_N\ep^2}
-\frac{r^2_2-r^2_1}{4G_N}\log\ep.
\ea
By comparing with the standard CFT result (\ref{partwocon}) using the bulk to boundary relation 
$Z_{CFT}=e^{iI_G}$, we obtain the central charge $c$ of the dual two dimensional CFT 
\ba
c=i\frac{3(r^2_2-r^2_1)}{4G_N}. \label{twodscent}
\ea

\subsubsection{$d=4$ Case}

For $d=4$, we can evaluate the on-shell action as follows:
\be
I_G=\frac{r^3_2-r^3_1}{G_N}\left[\frac{1}{48}r_{\infty}^3e^{3t_{\infty}}+\frac{3}{8}r_{\infty}^3e^{t_{\infty}}\right]
\ee
Via the relation (\ref{cutoffto}), we obtain
\be
I_G=\frac{r_{2}^3-r^3_1}{48G_N\ep^3}+\frac{3(r_2^3-r^3_1)}{16G_N\ep}.
\ee
Note that there is no logarithmic term as in odd dimensions there is no conformal anomaly.

\subsubsection{$d=5$ Case}

For $d=5$, we can estimate on-shell action as follows:
\be
I_G=\frac{\pi(r^4_2-r^4_1)}{16G_N}\left[\frac{1}{12}e^{4t_{\infty}}+\frac{2}{3}e^{2t_{\infty}}+2t_{\infty}\right].
\ee
In terms of the CFT cut off (\ref{cutoffto}), we find
\be
I_G=\frac{\pi (r_{2}^4-r_1^4)}{2G_N}\left[\f{1}{96\ep^4}+\frac{1}{12\ep^2}
-\frac{1}{4}\log\ep\right].
\ee
By comparing the logarithmic term in $Z_{CFT}=e^{iI_G}$ with (\ref{actionanofour}),
we can evaluate 
\be
a=-i\frac{\pi(r_{2}^4-r_1^4)}{32G_N}. \label{confouraa}
\ee

\subsection{Holographic entanglement entropy}
\begin{figure}
  \centering
  \includegraphics[width=8cm]{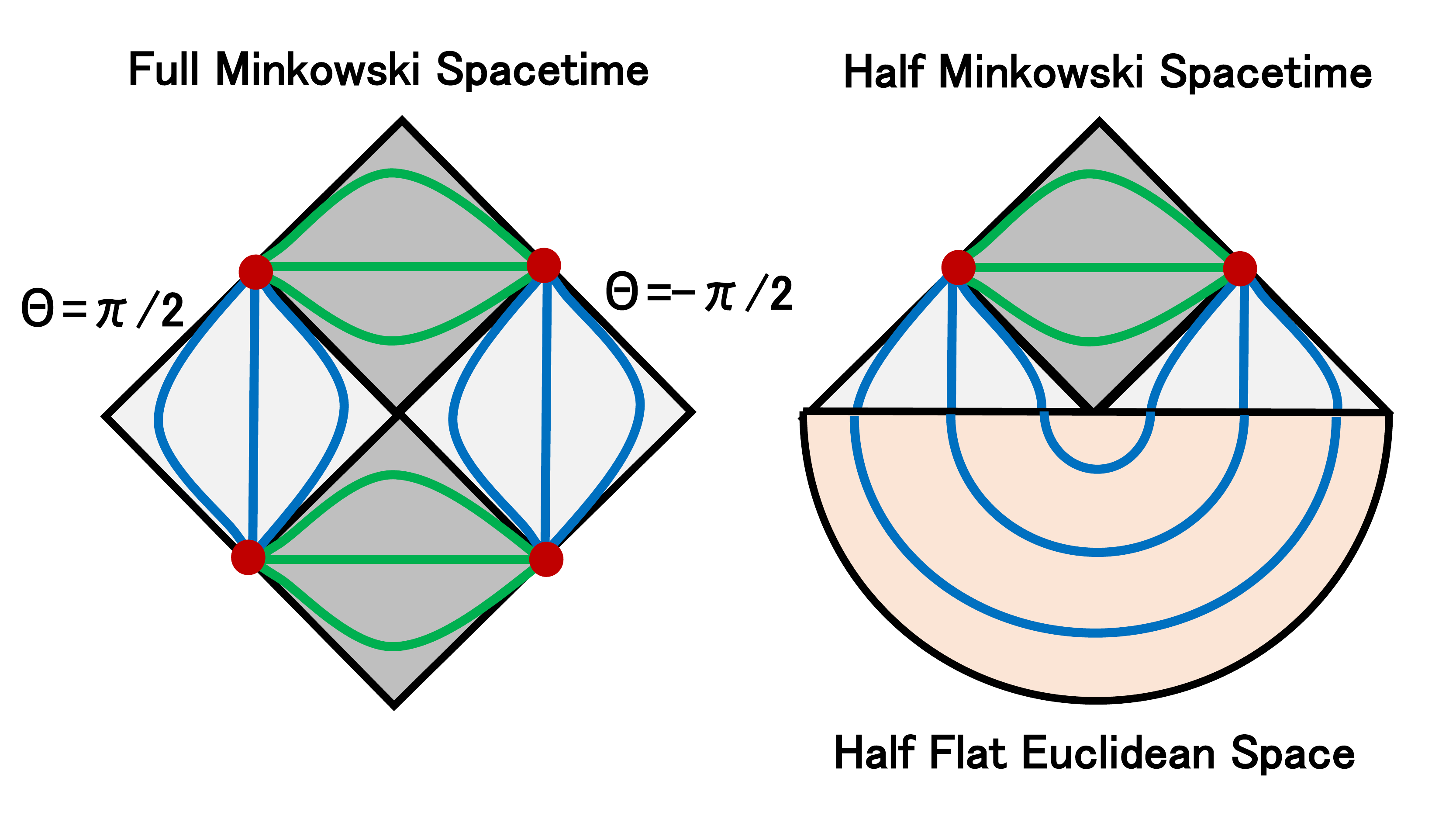}
  \caption{Sketches of the extremal surfaces which calculate the holographic entanglement entropy. We chose the maximal subsystem $A$ i.e. $\theta_0=\frac{\pi}{2}$. The left panel depicts the extremal surfaces in the full Minkowski spacetime. The right one sketches those in the spacetime which describes the Hartle-Hawking wave function, i.e. a half Euclidean flat space (past) plus a half Minkowski spacetime (future).
  The green and blue surfaces are the minimal surface in hyperbolic patch and the extremal surfaces in the de Sitter patch, respectively.}
\label{fig:extsur}
\end{figure}

The extremal surface which computes the holographic entanglement entropy (\ref{HEE}) can be constructed from a family of extremal surfaces in the de Sitter slice.
Thus, for a fixed value of $r$, it is given by the extremal surface $\gamma^{dS}_A$ in dS$_d$ calculated in Appendix \ref{appendix:ds}.
Consider the metric of dS$_d$ given by  
\ba
ds^2=-dt^2+\cosh^2t (d\theta_1^2+\sin^2\theta_1d\Omega_{d-2}).  \label{dssph}
\ea
The area of an extremal surface which stretches between $\theta=-\theta_0$ to $\theta=\theta_0$ on the sphere S$^{d-1}$ at the asymptotic boundary $t=t_\infty\to\infty$ is given by
\ba
A(\gamma^{dS}_A)=i\omega_{d-3}\cdot \int^{\infty}_{\frac{\ep}{L}}dy\frac{(1+y^2)^{\frac{d-4}{2}}}{y^{d-2}},
\ea
where $\ep$ and $L$ are related to the cut off $\rho_\infty$ and $\theta_0$ via (\ref{CUTaa}) and  (\ref{CUTbb}). Note that this extremal surface is time-like and extends to the other sphere S$^{d-1}$ at $t=-t_{\infty} \to -\infty$ instead of going back to the original sphere as is typical in the dS/CFT \cite{Hikida:2021ese,Hikida:2022ltr} (refer to the left panel of Fig.\ref{fig:extsur}). It is also possible to replace $t<0$ spacetime with a Euclidean flat space:
\ba
ds^2=dr^2+r^2(d\tau^2+\cos^2\tau d\Omega^2_{d-1}),  \label{Eflat}
\ea
by performing a Wick rotation $\tau=it$. This provides the Hartle-Hawking construction of the wave function of flat space (refer to the right panel of Fig.\ref{fig:extsur}). In this case we can connect the extremal surface inside the Euclidean space \cite{Hikida:2021ese,Hikida:2022ltr}. Motivated by this, we here compute the area of extremal surface for the half of Lorentzian dS$_d$ i.e. $t\geq 0$. Thus to recover the holographic entanglement entropy for full wedge $-t_\infty<t<t_\infty$, we can simply double the result as in the left panel of Fig.\ref{fig:extsur}. If we would like to consider the holographic entanglement entropy in the Hartle-Hawking state, then we need to add a Euclidean minimal surface area. In this paper we have in mind the former prescription.
 
For $d=3,4$ and $5$ we obtain the following results:
\ba
&& A(\gamma^{dS}_A)_{d=3}=2i\log\frac{2L}{\delta}=2i\log (e^{t_{\infty}}\sin\theta_0),\no
&&  A(\gamma^{dS}_A)_{d=4}=2i\pi\frac{L}{\delta}=\pi i\sin\theta_0 e^{t_\infty},\no
&& A(\gamma^{dS}_A)_{d=5}=4\pi i\left[\frac{L^2}{2\delta^2}
+\frac{1}{2}\log\frac{2L}{\delta}+\frac{1}{4}\right]\no
&& =\frac{\pi i}{2}\sin^2\theta_0 e^{2t_\infty}+2\pi i\log\left(\frac{\sin\theta_0}{4}e^{t_\infty}\right)
-\pi i\cos\theta_0(2+\cos\theta_0).
\ea
Thus the total area of extremal surface in $W^{ds}$ is given by 
\ba
A(\Gamma^{dS}_A)=\int^{r_2}_{r_1}r^{d-1}dr \cdot A(\gamma^{dS}_A)=\frac{1}{d-1}(r^{d-1}_2-r^{d-1}_1)
 A(\gamma^{dS}_A).   \label{minareadS}
\ea

In this way the final expression of the holographic entanglement entropy reads
\ba
S_A=\frac{A(\Gamma^{dS}_A)}{4G_N}=\frac{(r^{d-1}_2-r^{d-1}_1)}{4(d-1)G_N}
 A(\gamma^{dS}_A).\label{sadsx}
\ea
If we consider the Hartle-Hawking prescription of flat space (i.e. the right panel of Fig.\ref{fig:extsur}), we need to add the extra contribution from the extremal surface in Euclidean geometry, denoted by $S^{(E)}_A$. This is computed by setting $A(\gamma^{ds}_A)$ to be the area of $d-2$ dimensional semi-sphere in (\ref{sadsx}), which leads to
\ba
S^{(E)}_A=\frac{(r_2^{d-1}-r_1^{d-1})\omega_{d-1}}{8(d-1)G_N}.
\ea

For $d=3$, we can explicitly evaluate $S_A$ in (\ref{sadsx}) as follows
\ba
S_A=\frac{i(r_2^2-r_1^2)}{8G_N}\log\left(\frac{\sin^2\theta_0}{\ep^2}\right),
\ea
where we employed (\ref{cutoffto}). By comparing this with the standard formula 
(\ref{CCfor}),  we can read off the value of central charge $c$ of the dual two dimensional CFT:
\ba
c=\frac{3i}{4G_N}(r_2^2-r_1^2).  \label{centracds}
\ea
This agrees with the result (\ref{twodscent}) obtained from the partition function.

For $d=5$, we obtain
\ba
S_A=i\frac{\pi (r^4_2-r^4_1)}{16G_N}\left[\f{\sin^2\theta_0}{2\ep^2}+2
\log\left(\frac{\sin\theta_0}{4\ep}\right)+O(1)\right].
\ea
By comparing this with the standard formula in 4D CFT (\ref{EEano}),
we can read off the value of the central charge $a$:
\ba
a=-\frac{\pi i(r^4_2-r^4_1)}{32G_N}.  \label{centralab}
\ea
Indeed, this reproduces our previous estimation (\ref{confouraa}) from the partition function.

\subsection{Scalar Field Propagation}\label{sec:dsscalar}

Now we consider a free massive scalar field in our wedge geometry $W^{ds}$ defined by $r_1\leq r\leq r_2$. We again impose the Dirichlet or Neumann boundary condition on the boundary $r=r_{1,2}$. As we will show below, the spectrum of $\lambda_k$, where the dual operator dimension reads $\Delta=1+i\lambda_k$ consist of the infinitely many real values of $\lambda_k$ and a finite number of imaginary values of $\lambda_k$.  The presence of the former, where the conformal dimension (\ref{confghw}) gets complex valued, again implies that the dual CFT on S$^2$ is non-unitary, as in the dS/CFT correspondence. In the dS$_3/$CFT$_2$ duality, we find the formula for the conformal dimension $\Delta=1\pm\s{1-M^2}$ \cite{Strominger:2001pn}, where $M$ is the mass of scalar in the dS$_3$. If we interpret our wedge holography result in terms of dS$_3/$CFT$_2$, we find a finite number of scalar fields in the range $0<M<1$ and an infinite number of scalar fields with $M>1$. 

\subsubsection{Dirichlet boundary condition}
Using (\ref{dswave}) and (\ref{frti}), the Dirichlet boundary condition for the scalar reads
\ba
\ti{f}_p(r_1)=0,~\ti{f}_p(r_2)=0.
\label{ds_dir_bndy}
\ea
This is equivalent to find such values of $\nu=\s{1+p^2}$ which are solutions to
\ba
\label{dsDdef}
D^{ds}(\nu,x_1,x_2)=J_{\nu}(x_1)H^{(1)}_{\nu}(x_2)
-J_{\nu}(x_2)H^{(1)}_{\nu}(x_1)=0.
\ea

\begin{figure}[hhh]
  \centering
  \includegraphics[width=8cm]{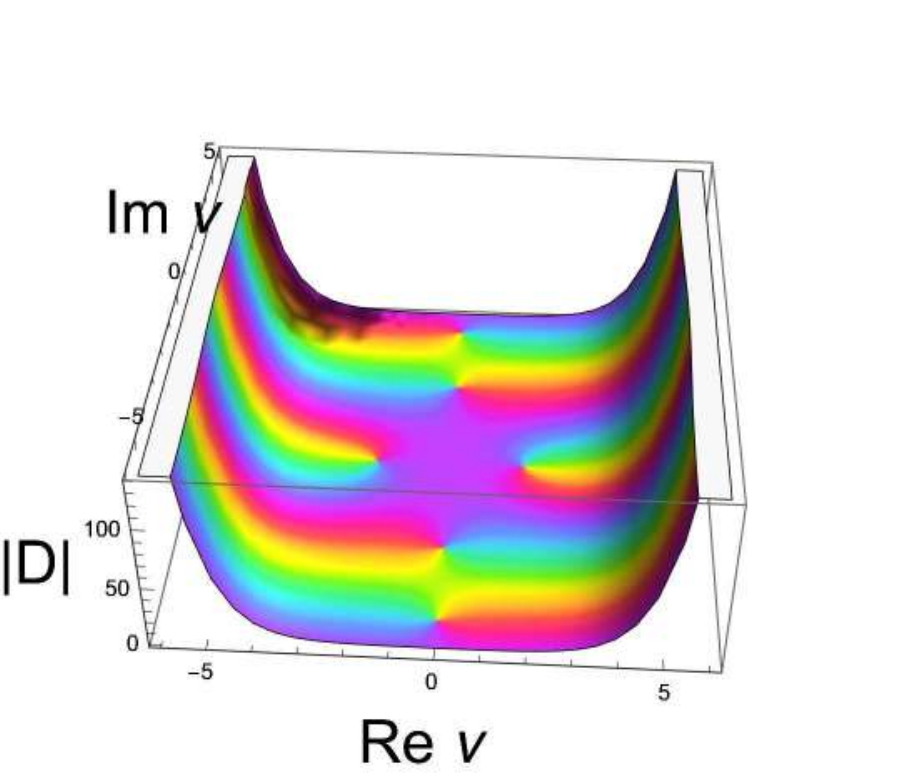}
   \includegraphics[width=6cm]{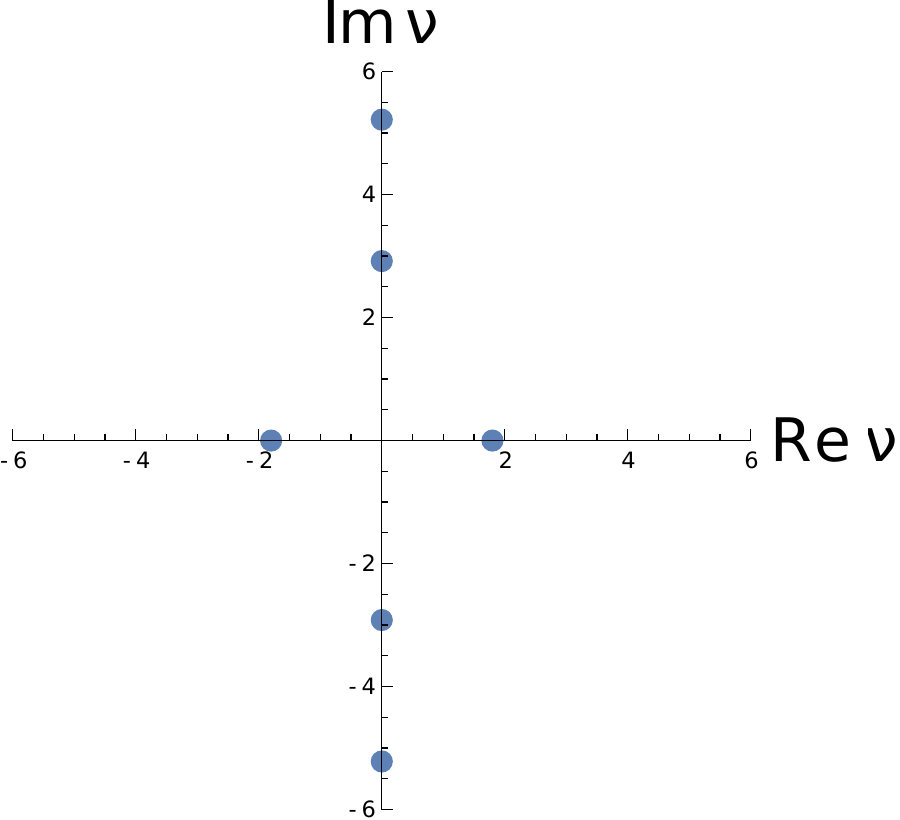}
  \caption{Plots of $|D^{dS}(\nu,x_1,x_2)|$ for $x_1=1$ and $x_2=5$ as a function of $\nu$ (left) and plots of zero point of $D^{dS}$ (right). The color of left figure represents $D$'s phase.}
\label{fig:ds_dir}
\end{figure}

By studying numerically, as plotted in Fig.\ref{fig:ds_dir}, we find that there is an infinite number of solutions for discrete imaginary values of $\nu$ together with
a finite number of solutions for real values of $\nu$. In appendix \ref{sec:ApD}, we analytically explain this behavior of solutions. We also find that the number of real valued solutions of $\nu$ increase as $r_2$ gets larger and the solutions with imaginary $\nu$ get dense in the limit $r_1\to 0$.

\subsubsection{Neumann boundary condition}

\begin{figure}[hhh]
  \centering
  \includegraphics[width=8cm]{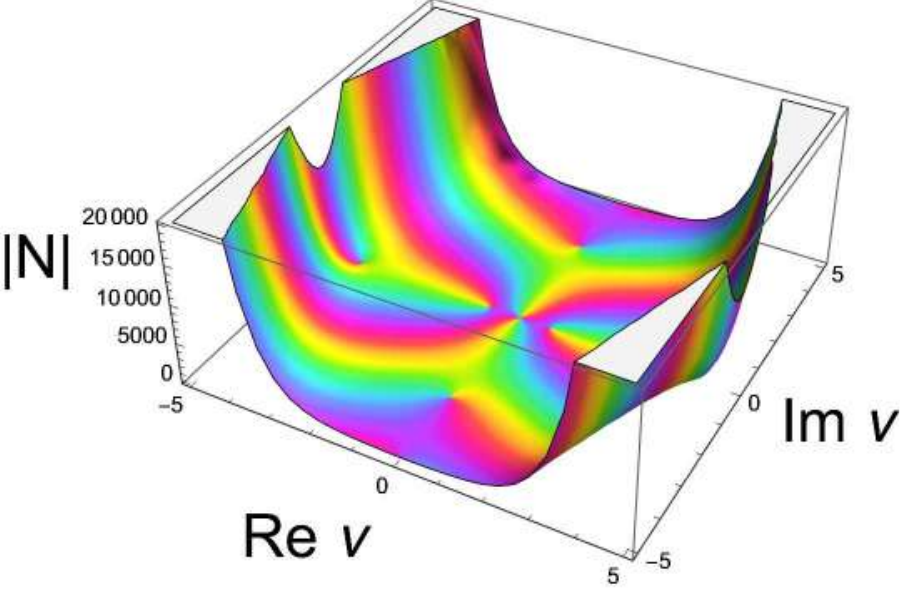}
   \includegraphics[width=6cm]{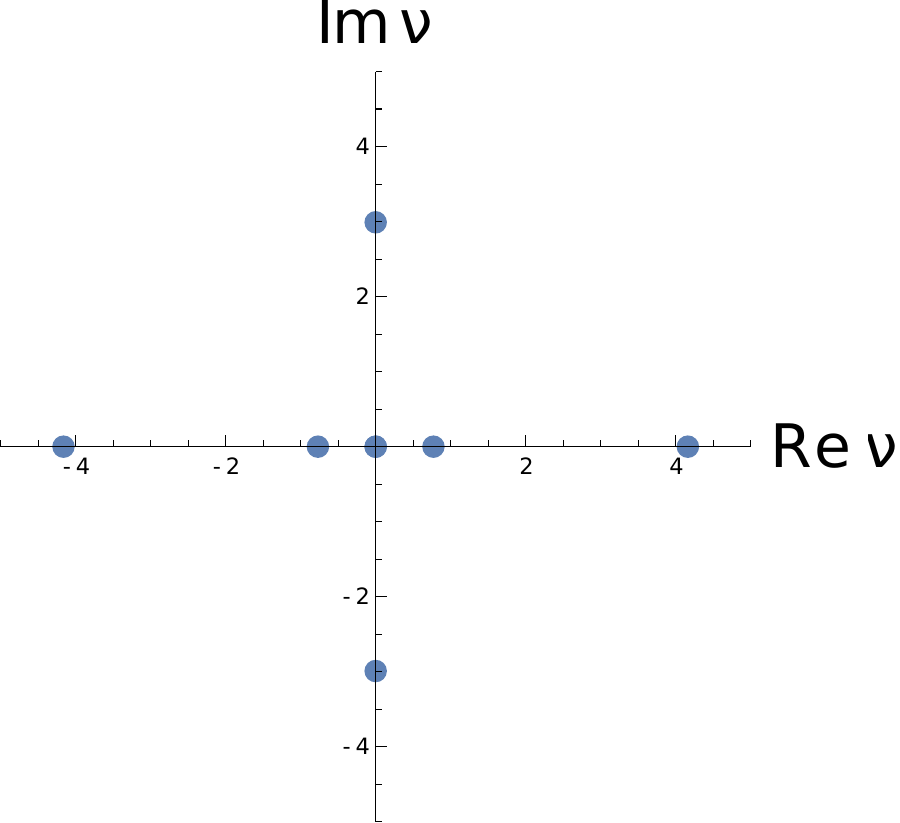}
  \caption{Plots of $|N^{ds}(\nu, x_1,x_2)|$ for $x_1=1$ and $x_2=5$ as a function of $\nu$ (left) and plots of zero points of 
  $N^{ds}$ (right). The color of left figure represents $N$'s phase.}
\label{fig:ds_neumann}
\end{figure}

Next we consider the case that we impose the Neumann boundary condition on the two EOW-branes:
\begin{equation}
\label{bcna}
\begin{aligned}
\partial_{r}\ti{f}_p(r_1)&=0,~
\partial_{r}\ti{f}_p(r_2)&=0,
\end{aligned}
\end{equation}
where the function $\ti{f}_p(r)$ was defined in (\ref{frti}).
By using the recurrence formula of modified Bessel function
\ba
&&\partial_x J_\nu(x)=-\frac{1}{2}\left(J_{\nu+1}(x)-J_{\nu-1}(x)\right),\ \ \ 
\partial_x H^{(1)}_\nu(x)=-\frac{1}{2}\left(H^{(1)}_{\nu+1}(x)-H^{(1)}_{\nu-1}(x)\right),\no
&&\frac{J_\nu(x)}{x}=\frac{1}{2\nu}\left(J_{\nu+1}(x)+J_{\nu-1}(x)\right),\ \ \ 
\frac{H^{(1)}_\nu(x)}{x}=\frac{1}{2\nu}\left(H^{(1)}_{\nu+1}(x)+H^{(1)}_{\nu-1}(x)\right),
\ea
we can write (\ref{bcna}) as follows:
\begin{align}
\ap\left( 
\frac{\nu+1}{2\nu}J_{\nu+1}(mr_a)\right.&-\left.\frac{\nu-1}{2\nu}J_{\nu-1}(mr_a)\right)
+m\beta\left(
\frac{\nu+1}{2\nu}H^{(1)}_{\nu+1}(mr_a)-\frac{\nu-1}{2\nu}H^{(1)}_{\nu-1}(mr_a)
\right)=0,\nonumber
\end{align}
where $a=0,1$. This is equivalent to the search of values of $\nu=\s{1+p^2}$ which satisfy 
\ba
\label{dsNdef}
N^{ds}(\nu,x_1,x_2)&=&\left\{(\nu+1)J_{\nu+1}(x_1)-(\nu-1)J_{\nu-1}(x_1)\right\}\left\{(\nu+1)H^{(1)}_{\nu+1}(x_2)-(\nu-1)H^{(1)}_{\nu-1}(x_2)\right\} \no
&&-\left\{(\nu+1)J_{\nu+1}(x_2)-(\nu-1)J_{\nu-1}(x_2)\right\}\left\{(\nu+1)H^{(1)}_{\nu+1}(x_1)-(\nu-1)H^{(1)}_{\nu-1}(x_1)\right\} \no
&=&0.
\ea
where $x_{1,2}= m r_{1,2}$. By studying numerically, as plotted in Fig.\ref{fig:ds_neumann}, we find that there is an infinite number of solutions for discrete values of $\nu$ together with
a finite number of solutions for real values of $\nu$.  The properties of the solutions $\nu$ are similar to the Dirichlet case. Refer to appendix \ref{sec:ApN} for more details.

\subsubsection{Two point function}
We can evaluate the two point functions as we did for the wedge holography in the hyperbolic patch in section \ref{sec:Hscalar}, by using the scalar field profile (\ref{bdybeads}). The result is identical to (\ref{correlationf}), expect that there are two spheres in future and past. If we call the operator inserted on the future and past sphere $O_p^{(+)}$ and $O_p^{(-)}$, respectively, then the two point functions read
\ba
&& \la O^{(\pm)}_p(\theta_1,\vp_1)O^{(\pm)}_p(\theta_2,\vp_2)\lb\propto \left(1-\cos\gamma_{12}\right)^{-\Delta},\no
&& \la O^{(\pm)}_p(\theta_1,\vp_1)O^{(\mp)}_p(\theta_2,\vp_2)\lb\propto \left(1+\cos\gamma_{12}\right)^{-\Delta},
\label{correlationfds}
\ea
where $\gamma_{12}$ was given in (\ref{gammadexf}). This means that an operator inserted at a point on the future sphere is equivalent to that inserted at its antipodal point on the past sphere. Under this identification, the two point functions agree with the CFT expectation.

\section{Is celestial holography a wedge holography ?}

In the previous sections, we present two new setups of wedge holography in flat spacetime: hyperbolic slices and de Sitter slices, as explained in section \ref{sec:setup} and depicted in Fig.\ref{fig:wedgeah}. In this section we would like to combine these two as in Fig.\ref{fig:setupp} to approach the celestial holography, which argues that $d+1$ dimensional gravity in a full Minkowski spacetime is dual to a CFT on the celestial sphere S$^{d-1}$. As we will see below, as long as we consider the vacuum configurations of celestial holography, it fits nicely with the wedge holography. However, if we consider excitations in celestial holography by gravitational waves, we will see that we need to modify boundary conditions of the flat space wedge holography we considered in previous section. 

\subsection{Partition function in Minkowski Spacetime}

Let us first calculate the partition function of celestial holography in Minkowski spacetime by regarding the on-shell gravity action as the CFT free energy simply by extending the standard bulk-boundary relation \cite{Gubser:1998bc,Witten:1998qj} of AdS/CFT.
We take the range of $\eta$ and $r$ to be (\ref{cutoffb}). Then we can simply add up the on-shell actions (\ref{Haction}) and (\ref{dSaction}) in the wedge holography by setting 
\ba
\eta_2=\eta_\infty,\ \ \ r_2=r_\infty,\ \ \ \eta_1=r_1=0.  \label{indetifgys}
\ea
This leads to
\be
I_G=\frac{1}{4\pi G_N}r_{\infty}^{d-1}\omega_{d-1}I_{d-1}-\frac{1}{4\pi G_N}\eta_{\infty}^{d-1}\omega_{d-1}J_{d-1}.
\ee
Here we doubled the result to cover the full Minkowski space i.e. not only $t>0$ but also $t<0$. This is evaluated in each dimension explicitly. For example, $d=3,4,5$ we obtain
\ba
&& d=3:\ \ \ I_G= \frac{r^2_\infty-\eta^2_\infty}{8G_N\ep^2}
-\frac{r^2_\infty+\eta^2_\infty}{2G_N}\log\ep,\no
&& d=4:\ \ \ I_G=\frac{r_{\infty}^3-\eta_{\infty}^3}{24G_N\ep^3}+\frac{3(r_{\infty}^3+\eta_{\infty}^3)}{8G_N\ep},\no
&& d=5:\ \ \ I_G=\frac{\pi(r_{\infty}^4-\eta_{\infty}^4)}{96G_N\ep^4}+\frac{\pi(r_{\infty}^4+\eta_{\infty}^4)}{12G_N\ep^2}-\frac{\pi(r_{\infty}^4-\eta_{\infty}^4)}{4G_N}\log\ep,
\ea
where $\ep$ is the UV cut off  such that $e^{-\rho_\infty}=e^{-t_\infty}=\ep$.
From the logarithmic terms, we can also read off the values of the central charges $c$ in $d=3$ and $a$ in $d=5$ as follows:
\ba
&& d=3:\ \ c=i\frac{3(r^2_\infty+\eta^2_\infty)}{2G_N},\no
&& d=5:\ \ a=i\frac{\pi(\eta_{\infty}^4-r_{\infty}^4)}{16G_N}. \label{centall}
\ea
These are consistent with standard behaviors in CFTs except that the central charges take imaginary values which show that the dual CFT is non-unitary. In our limit $\eta_\infty\to\infty$ and $r_\infty\to\infty$, the two dimensional CFT central charge becomes $c\to i\infty.$  Such a divergent central charge in the dual CFT has also been argued in \cite{Cheung:2016iub,Pasterski:2022lsl}. Moreover, it is intriguing to note that we can have $a=0$ for the central charge of the four dimensional CFT if we tune $\eta_\infty=r_\infty$.

\subsection{Holographic entanglement entropy in Minkowski spacetime}

We can calculate the holographic entanglement entropy in celestial holography in Minkowski spacetime. As before, we chose the subsystem $A$ to be $|\theta|\leq \theta_0$ on S$^{d-1}$.  For this, we add the contribution in hyperbolic patch (\ref{minareaH}) and the de Sitter patch (\ref{minareadS}) of the wedge holography by taking the range (\ref{indetifgys}) and double it to cover the entire spacetime. This leads to the total expression:
\ba
S_A=\frac{A(\Gamma^{H}_A)+A(\Gamma^{dS}_A)}{4G_N}=\frac{i\eta^{d-1}_\infty}{2(d-1)G_N}
 A(\gamma^{H}_A)+\frac{r^{d-1}_\infty}{2(d-1)G_N}
 A(\gamma^{dS}_A).
\ea

For example, we obtain explicit results for $d=3$ and $d=5$ as follows:
\ba
&& d=3:\ \ 
S_A=\frac{i}{4G_N}(\eta_\infty^2+r_\infty^2)\cdot \log\frac{\sin^2\theta_0}{\ep^2},\no
&& d=5:\ \ S_A=i\frac{\pi(\eta^4_\infty+r^4_\infty)}{16G_N\ep^2}+i\frac{\pi}{4G_N}
(r^4_\infty-\eta^4_\infty)\log\left(\frac{\sin\theta_0}{4\ep}\right)+O(1),
\ea
where $\ep$ is the UV cut off.
By comparing this with the general expressions (\ref{CCfor}) and (\ref{EEano}), we find the same central charges $c$ and $a$ which we obtained from the on-shell action in (\ref{centall}).

\subsection{Celestial holography versus wedge holography with excitations}

The celestial holography \cite{Pasterski:2016qvg,Raclariu:2021zjz} argues that four dimensional gravity on the Minkowski spacetime is dual to a two dimensional CFT on the celestial sphere S$^2$ at null infinity. One basic relation in the celestial holography is the connection between scattering amplitudes ${\cal A}(k_1,k_2,\ddd,k_N)$ of $N$ particles in four dimensions and correlation functions $\la O_1O_2\ddd O_N\lb_{S^2}$ of $N$ primary operators.  
For a scalar field dual to a scalar operator $O_\Delta$ with the dimension $\Delta$, this is explicitly written as follows
\ba
&&\la O_{\Delta_1}(\theta_1,\vp_1)O_{\Delta_2}(\theta_2,\vp_2)\ddd O_{\Delta_N}(\theta_N,\vp_N)\lb_{S^2}\no
&& =\left[\prod_{i=1}^{N}\int dX^\mu_i \phi^{\Delta_i,(\pm)}(X^\mu_i,\theta_i,\vp_i)\int dk^\mu_i e^{ik^\mu_iX_{\mu i}}\right]{\cal{A}}(k_1,k_2,\ddd,k_N).
\ea
In this correspondence, the functions $\phi^{\Delta_i,(\pm)}$ are called conformal primary wave functions. The superscript $(+)$ and $(-)$ correspond to out-going and in-coming particle, respectively. They are explicitly given by the following expression \cite{Pasterski:2016qvg}:
\ba
&& \phi^{\Delta,\pm}(X^\mu,\theta_0,\vp_0)\equiv \frac{\left(\s{-X^\mu X_\mu}\right)^{\Delta-1}}{(q^\mu X^{\pm}_\mu)^\Delta}K_{\Delta-1}\left(m\s{X^\mu X_\mu}\right).
\label{celwf}
\ea
Here $X^\mu$ is the four dimensional Minkowski coordinate, which is related to the hyperbolic patch coordinate and de Sitter patch one via
\ba
&& (X^0,X^1,X^2,X^3)|_{X^\mu X_\mu<0}=\eta\cdot(\cosh\rho,\sinh\rho \sin\theta\cos\phi,\sinh\rho \sin\theta\sin\phi,\sinh\rho \cos\theta),\no
&& (X^0,X^1,X^2,X^3)|_{X^\mu X_\mu>0}=r\cdot(\sinh t,\cosh t \sin\theta\cos\phi,\cosh t \sin\theta\sin\phi,\cosh t \cos\theta).\nonumber
\ea
and $q^\mu$ is the null vector 
\ba
(q^0,q^1,q^2,q^3)=\frac{2}{1+\cos\theta_0}\cdot (1,\sin\theta_0\cos\phi_0,\sin\theta_0\sin\phi_0,\cos\theta_0),
\ea
which specifies the direction of particle on the celestial sphere.
We also introduced  the $i\ep$ regularization $X^{\mu,\pm}=X^\mu\pm i\ep\{-1,0,0,0\}$. 

This wave function $\phi^{\Delta_i,(\pm)}$ can be interpreted as a point-like excitation on the celestial sphere due to the out-going or in-coming wave. In terms of hyperbolic/de Sitter patch coordinate, the conformal primary wave functions (\ref{celwf}) read (setting $\ep=0$)
\ba
&& \phi^{\Delta,\pm}(X^\mu,\theta_0,\vp_0)|_{X^\mu X_\mu<0}=\frac{K_{\Delta-1}(m\eta)}{\eta} \left(\frac{1+\cos\theta_0}{2}\right)^\Delta \left(\cosh\rho-\cos\gamma\sinh\rho\right)^{-\Delta}, \no
\label{hswave}\\
&& \phi^{\Delta,\pm}(X^\mu,\theta_0,\vp_0)|_{X^\mu X_\mu>0}=-\frac{\pi i}{2}\frac{H^{(a)}_{\Delta-1}(mr)}{r} \left(\frac{1+\cos\theta_0}{2}\right)^\Delta \left(\sinh t-\cos\gamma\cosh t\right)^{-\Delta},\no\label{dswavee}
\ea
where the type of the Hankel function $a=1,2$ corresponds to the out-going $(+)$ and in-coming $(-)$ wave. 
Indeed, these are among the class of the scalar field solutions  (\ref{phiH}) and (\ref{phidS}) with a delta-functional source on the celestial sphere S$^2$. In the hyperbolic patch, the celestial holography and our wedge holography discussed in section \ref{sec:HP} have the same boundary condition for a massive free scalar i.e. the Dirichlet (or Neumann) boundary condition\footnote{Note that in the UV limit $\eta_\infty\to \infty$ of celestial holography, the Dirichlet and Neumann boundary condition at $\eta=\eta_\infty$ for the scalar field are identical. Thus we can consider this the Neumann boundary condition.} at $\eta=\eta_\infty$. This is clear from the expression (\ref{dswavee}) as the Bessel function $K_\nu$ appears, which exponentially decays as $K_\nu(z)\sim e^{-z}$ for large $|z|$.

However, the boundary condition we impose in the $r$ direction of the de Sitter patch looks different between the celestial holography and our wedge holography. In the former, as in (\ref{dswavee}), we impose the out-going or in-coming boundary condition at $r=r_\infty$, while in the latter we require the Dirichlet (or Neumann) boundary condition. A similar observation is true for the gravitational wave mode, where we impose the out-going or in-coming boundary condition in celestial holography and we do the Neumann boundary condition (\ref{NBY}) in our wedge holography.
In this sense, if we want to interpret the celestial holography in terms of a wedge holography in flat space, we need to modify the boundary condition in the de Sitter patch at $r=r_\infty$. However, notice that in the computation of correlation functions, this difference of $r$ dependence only appears in the overall constant and thus does not affect the dependence of celestial sphere coordinate e.g. in (\ref{correlationfds}). 

Here, we should also notice that the conformal dimension $\Delta$, available in both hyperbolic patch and de Sitter patch, is $1+i\lambda$ where $\lambda$ is an arbitrary real value (see section \ref{sec:Hscalar}, \ref{sec:dsscalar}, Appendix \ref{ApD}). This result from our wedge holography is consistent with the principle series in celestial holography, which is constrained from "normalizable condition \cite{Pasterski:2017kqt}", not from boundary condition.

\section{Conclusions and discussions}

In this paper, we proposed extensions of wedge holography to a flat spacetime, largely motivated by the recent developments of celestial holography. A wedge holography \cite{Akal:2020wfl} is in general a codimension two holographic duality between a gravitational theory in a wedge region and a CFT on its tip. 

As the first example of wedge holography in a flat spacetime, we argued that a $d+1$ dimensional region surrounded by two $d$ dimensional hyperbolic spaces (depicted in the left panel of Fig.\ref{fig:wedgeah}) is dual to a non-unitary CFT on S$^{d-1}$. We imposed the Neumann boundary condition (\ref{NBY}) for gravitational modes on the two boundaries i.e. the end of the world-branes (EOW brane). We calculated the on-shell gravity action, holographic entanglement entropy and two point functions in the gravity dual and found that they agree with general expectations in CFTs. The superrotation symmetry at each hyperbolic slice explains the conformal symmetry of the dual Euclidean CFT.

In this example, it is intriguing that a time-like direction in addition to a space-like radial direction emerges from the Euclidean CFT. We found that the central charges in even dimensional CFTs dual to the wedge region take imaginary values and also that the conformal dimensions dual to a bulk scalar become complex valued. These two unusual properties show that the dual CFT is non-unitary.
This is not at all surprising because there is a good reason to believe that the holographic duality where a real time direction emerges involves non-unitary theory, as is expected in the dS/CFT duality \cite{Strominger:2001pn,Maldacena:2002vr}. Indeed, the known CFT duals of dS/CFT in four dimensions \cite{Anninos:2011ui} and in three dimensions \cite{Hikida:2021ese} are all non-unitary. It will be interesting to explore this wedge holography from more sophisticated viewpoints such as higher point functions, entanglement wedges and various excited states.

The second example of flat space wedge holography, which we proposed in this paper, is for gravity in the $d+1$ dimensional wedge region (the right panel of Fig.\ref{fig:wedgeah}) bounded by two $d$ dimensional de Sitter spaces. We again impose the Neumann boundary condition (\ref{NBY}) on the two EOW branes. We evaluated the on-shell gravity action, holographic entanglement entropy and two point functions in the gravity dual and again confirmed that they are consistent with general expectations in CFTs.  The superrotation symmetry at each de Sitter slice explains the conformal symmetry of the dual Euclidean CFT. This wedge holography can be regarded as a slightly 'fatten' version of dS/CFT correspondence by simply adding a spacial interval. Therefore our calculations and results were parallel with that in dS/CFT. Indeed, the central charges in even dimensional CFTs on S$^{d-1}$ tuned out to take imaginary values. We found that there are infinitely many scalar operators dual to a bulk scalar which have imaginary valued conformal dimensions. In addition there are a finite number of scalar operators with real valued conformal dimensions. 

Since the full Minkowski spacetime can be regarded as a union of the hyperbolic patch and de Sitter patch, we finally considered a possibility that the celestial holography for the former can be interpreted as a combination of the hyperbolic and de Sitter sliced wedge holography.
We found that the results of the on-shell action and holographic entanglement for the flat Minkowski spacetime, which are simply the sum of those in hyperbolic and de Sitter sliced wedge holography, look consistent with the CFT expectations. However, if we consider excitations such as the bulk scalar field, we found that the wedge holography in the de Sitter patch has a different boundary condition than that in the celestial holography. The former is either Dirichlet or Neumann and the latter is out-going or in-coming. On the other hand, 
in the hyperbolic patch, our wedge holography and celestial holography assume the same boundary condition.
Therefore, we need to modify the usual boundary condition of wedge holography, which is Neumann (\ref{NBY}) for metric perturbation modes, 
to the out-going or in-coming boundary condition in order to interpret the celestial holography as a wedge holography. 

It would be an intriguing future direction to explore more the fundamental mechanism of celestial holography and generalize the flat space holography to non-trivial geometries such as Schwarzschild black holes.

\section*{Acknowledgements}

We are grateful to Ibrahim Akal, Taishi Kawamoto, Sinji Mukohyama, Hidetoshi Omiya, Shan-Ming Ruan, Yu-ki Suzuki, Yusuke Taki, Tomonori Ugajin and Zixia Wei, for useful discussions. We thank very much Andrew Strominger for valuable comments. This work is supported by the Simons Foundation through the ``It from Qubit'' collaboration
and by MEXT KAKENHI Grant-in-Aid for Transformative Research Areas (A) through the ``Extreme Universe'' collaboration: Grant Number 21H05182 and 21H05187.
This work is also supported by Inamori Research Institute for Science and World Premier International Research Center Initiative (WPI Initiative)
from the Japan Ministry of Education, Culture, Sports, Science and Technology (MEXT),
by JSPS Grant-in-Aid for Scientific Research (A) No.~21H04469 and
by JSPS Grant-in-Aid for Challenging Research (Exploratory) 18K18766.

\appendix

\section{Useful identities of Legendre functions}

The associated Legendre function is defined by (we follow \cite{Book})
\ba
&& P^\mu_\nu(z)=\frac{1}{\Gamma(1-\mu)}\left(\frac{z+1}{z-1}\right)^{\mu/2}
{}_2F_1\left(-\nu,\nu+1;1-\mu;\frac{1-z}{2}\right).\no
&& Q^\mu_\nu(z)=\frac{e^{\pi\mu i}\s{\pi}\Gamma(\mu+\nu+1)}{2^{\nu+1}\Gamma(\nu+3/2)} 
z^{-\mu-\nu-1}(z^2-1)^{\mu/2}
{}_2F_1\left(\frac{\mu+\nu+2}{2},
\frac{\mu+\nu+1}{2};\nu+\frac{3}{2};\frac{1}{z^2}\right).\no
\ea
It is useful to note the asymptotic behavior in the $|z|\to\infty$
\ba
Q^\mu_\nu(z)\simeq e^{\mu\pi i}\s{\pi}\frac{\Gamma(\nu+\mu+1)}{\Gamma(\nu+3/2)(2z)^{\nu+1}},
\ea
and $z\to 1$
\ba
Q^\mu_\nu(z)\simeq \frac{e^{\pi \mu i}}{2}\left[\Gamma(\mu)\left(\frac{2}{z-1}\right)^{\mu/2}
+\frac{\Gamma(-\mu)\Gamma(\nu+\mu+1)}{\Gamma(\nu-\mu+1)}\left(\f{z-1}{2}\right)^{\mu/2}\right].
\label{expone}
\ea

The spherical harmonic function is defined by
\ba
 Y_{lm}(\theta,\phi)=(-1)^m\s{\frac{(2l+1)(l-m)!}{4\pi (l+m)!}}P^m_l(\cos\theta)e^{im\phi}. \label{spharmo}
 \ea
It satisfies the orthonormal condition:
\ba
\int^{\pi}_{0} d\theta\sin\theta\int^{2\pi}_0d\phi Y^*_{lm}(\theta,\phi)Y_{l'm'}(\theta,\phi)=\delta_{ll'}\delta_{mm'}.
\ea
We can also show
\ba
\sum_{l=0}^\infty\sum_{m=-l}^l Y^*_{lm}(\theta,\phi)Y_{lm}(\theta_0,\phi_0)
=\frac{1}{\sin\theta}\delta(\theta-\theta_0)\delta(\phi-\phi_0)\equiv\delta^2(\Omega-\Omega_0).
\ea

The additivity theorem is also useful:
\ba
Y_{l,0}(\gamma)=\s{\frac{4\pi}{2l+1}}\sum_{m=-l}^lY^*_{lm}(\theta,\phi)Y_{lm}(\theta_0,\phi_0), \label{additive}
\ea
where $\gamma$ is defined by (\ref{gammadef}).

The following integral formula is also useful (this is eq.$7.228$ of \cite{Book})
\ba
\int^{1}_{-1}dx \frac{P_{n}(x)}{(z-x)^{\mu+1}}=\frac{2}{\Gamma(1+\mu)}(z^2-1)^{-\mu/2}e^{-i\pi\mu}Q^\mu_n(z).  \label{intfora}
\ea
In particular by taking the limit $z=1$ we obtain
\ba
\int^{1}_{-1}dx \frac{P_{n}(x)}{(1-x)^{\mu+1}}=(-1)^n\frac{2^{-\mu}\Gamma(-\mu)^2}{\Gamma(n-\mu+1)\Gamma(-\mu-n)}.\label{exptwo}
\ea

\section{Minimal surfaces and geodesic length in H$_d$}\label{appendix:hdmin}
Here we summarize minimal surfaces and geodesic length in the hyperbolic space H$_d$.

\subsection{Minimal surfaces}
Consider $H_d$ whose metric is given by (\ref{hypbl}).
This is described by a coordinate $(X_0,X_1,\ddd,X_d)$ on the surface
\ba
X_0^2=X_1^2+\ddd+X_d^2+1,
\ea
in $R^{1,d}$, via the coordinate transformation: 
\ba
&& X_0=\cosh\rho,\no
&& X_1=\sinh\rho\cos\theta_1,\no
&& X_2=\sinh\rho\sin\theta_1\cos\theta_2,\no
&& \ddd  ,\no
&& X_d=\sinh\rho\sin\theta_1\sin\theta_2\ddd \sin\theta_{d-1}.
\ea

We can also map this to the Poincare coordinate as
\ba
&& X_0=\frac{z}{2}
\left(1+\frac{x^2+1}{z^2}\right),\no
&& X_1=-\frac{z}{2}
\left(1-\frac{1-x^2}{z^2}\right),\no
&& X_i=\frac{x_{i-1}}{z} \ \ \ (i=2,3,\ddd,d),\no
\ea
leading to the metric
\ba
ds^2=\frac{dz^2+dx^2_1+\ddd+dx^2_{d-1}}{z^2}.  \label{Poic}
\ea

It is well-known that a class of minimal surfaces in (\ref{Poic}) is given by $d-2$ dimensional semi-spheres.
\ba
&& x_{d-1}=0,\no
&& z^2+x_1^2+\ddd+x_{d-2}^2=L^2.
\ea
In terms of the original coordinate (\ref{hypbl}) of $H_d$, this is expressed as 
\ba
&&1+\sinh^2\rho\sin^2\theta=L^2(\cosh\rho+\sinh\rho \cos\theta_1)^2,\no
&& \theta_{d-1}=0,
\ea
while the angles $(\theta_2,\ddd,\theta_{d-2})$ are free.
We introduce $\theta_0$ such that we have $\theta=\pm\theta_0$ at the boundary $\rho=\rho_\infty\to \infty$.
This is given by 
\ba
L=\frac{\sin\theta_0}{1+\cos\theta_0}. \label{CUTa}
\ea
Note also that the cut off in the Poincare coordinate $z=\ep$ is mapped into that in the original coordinate as
\ba
\frac{1}{\delta}=\left(\frac{1+\cos\theta_0}{2}\right)e^{\rho_\infty}+\left(\frac{1-\cos\theta_0}{2}\right)e^{-\rho_\infty}.  \label{CUTb}
\ea

\subsection{Geodesic length}

If we consider two points $P_1$ and $P_2$
\ba
&& P_1=(\rho_1,\theta^{(1)},\Omega_{d-2}),\no
&& P_2=(\rho_2,\theta^{(2)},\Omega_{d-2}).
\ea
the geodesic distance $D_{12}$ in the hyperbolic space H$_d$ reads 
\ba
\cosh D_{12}=\cosh\rho_1\cosh\rho_2-\sinh\rho_1\sinh\rho_2\cos(\theta^{(1)}-\theta^{(2)}).
\ea
In the limit $\rho_1=\rho_2=\rho_\infty$, this leads to
\ba
D_{12}
&=& 2\rho_{\infty}+\log\left(\sin^2\frac{\theta_1-\theta_2}{2}\right).\label{geoaa}
\ea

The geodesic is explicitly given by
\ba
\tan\left[\theta-\frac{\theta^{(1)}+\theta^{(2)}}{2}\right]=\frac{1}{\cosh\rho}\s{\frac{\sinh^2\rho}{\sinh^2\rho_*}-1},
\ea
where we set
\ba
\tan\left[\frac{\theta^{(1)}-\theta^{(2)}}{2}\right]=\frac{1}{\sinh\rho_*}.
\ea

\section{Extreme surfaces and geodesic length in dS$_d$}\label{appendix:ds}
Here we summarize minimal surfaces and geodesic length in the de Sitter spacetime dS$_d$.

\subsection{Extremal surfaces}

Consider dS$_d$ whose metric is given by (\ref{dssph}).
This is described by a coordinate $(X_0,X_1,\ddd,X_d)$ on the surface
\ba
X_0^2+1=X_1^2+\ddd+X_d^2,
\ea
in $R^{1,d}$, via the coordinate transformation: 
\ba
&& X_0=\sinh t,\no
&& X_1=\cosh t\cos\theta_1,\no
&& X_2=\cosh t\sin\theta_1\cos\theta_2,\no
&& \ddd  ,\no
&& X_d=\cosh t\sin\theta_1\sin\theta_2\ddd \sin\theta_{d-1}.
\ea

We can also map this to the Poincare coordinate as
\ba
&& X_0=-\frac{z}{2}
\left(1-\frac{x^2+1}{z^2}\right),\no
&& X_1=\frac{z}{2}
\left(1+\frac{1-x^2}{z^2}\right),\no
&& X_i=\frac{x_{i-1}}{z} \ \ \ (i=2,3,\ddd,d),\no
\ea
leading to the metric
\ba
ds^2=\frac{-dz^2+dx^2_1+\ddd+dx^2_{d-1}}{z^2}.  \label{Poicc}
\ea

It is well-known that a class of minimal surfaces in (\ref{Poicc}) is given by $d-2$ dimensional semi-spheres.
\ba
&& x_{d-1}=0,\no
&& x_1^2+\ddd+x_{d-2}^2=z^2+L^2.
\ea
In terms of the original coordinate (\ref{dssph}) of dS$_d$, this is expressed as 
\ba
&&\cosh^2t\sin^2\theta=L^2(\sinh t +\cosh t \cos\theta_1)^2+1,\no
&& \theta_{d-1}=0,
\ea
while the angles $(\theta_2,\ddd,\theta_{d-2})$ are free.
We introduce $\theta_0$ such that we have $\theta=\pm\theta_0$ at the boundary $\rho=\rho_\infty\to \infty$.
This is given by 
\ba
L=\frac{\sin\theta_0}{1+\cos\theta_0}. \label{CUTaa}
\ea
Note also that the cut off in the Poincare coordinate $z=\delta$ is mapped into that in the original coordinate as
\ba
\frac{1}{\delta}=\left(\frac{1+\cos\theta_0}{2}\right)e^{t_\infty}-\left(\frac{1-\cos\theta_0}{2}\right)e^{-t_\infty}.  \label{CUTbb}
\ea

\subsection{Geodesic length}

If we choose two points $P_1$ and $P_2$ on dS$_d$: 
\ba
&& P_1=(t_1,\theta^{(1)},\Omega_{d-2}),\no
&& P_2=(t_2,\theta^{(2)},\Omega_{d-2}),
\ea
where we took the locations on $S^{d-2}$ are the same without losing generality owing to the $SO(d-1)$ symmetry.  The geodesic distance between 
$P_1$ and $P_2$, denoted by $D_{12}$, can be found as 
\ba
\cos D_{12}=\cos(\theta^{(1)}-\theta^{(2)})\cosh t_1\cosh t_2-\sinh t_1\sinh t_2.
\ea

If we choose $t_1=t_2=t_{\infty}\to \infty$, we find 
\ba
D_{12}\simeq 2it_{\infty}+i\log\left[\sin^2\left(\frac{\theta^{(1)}-\theta^{(2)}}{2}\right)\right]+\pi. \label{dSga}
\ea
The imaginary divergent contribution comes from the time-like geodesic and
the final real part $\pi$ does from the geodesic in an Euclidean space ($d$ dim. half sphere). For more detail of this and an interpretation in dS/CFT, refer to Fig.5 of \cite{Hikida:2022ltr}.

On the other hand, if we choose $t_1=-t_2=t_{\infty}\to \infty$, we obtain
\ba
D_{12}\simeq 2it_{\infty}+i\log\left[\cos^2\left(\frac{\theta^{(1)}-\theta^{(2)}}{2}\right)\right]. \label{dSgb}
\ea
Note that if we replace $\theta_2$ with the antipodal one $\theta_2+\pi$, then we get the behavior of (\ref{dSga}).

\section{Scalar field modes in de Sitter sliced wedges} \label{ApD}
\begin{figure}[hhh]
  \centering
  \includegraphics[width=7cm]{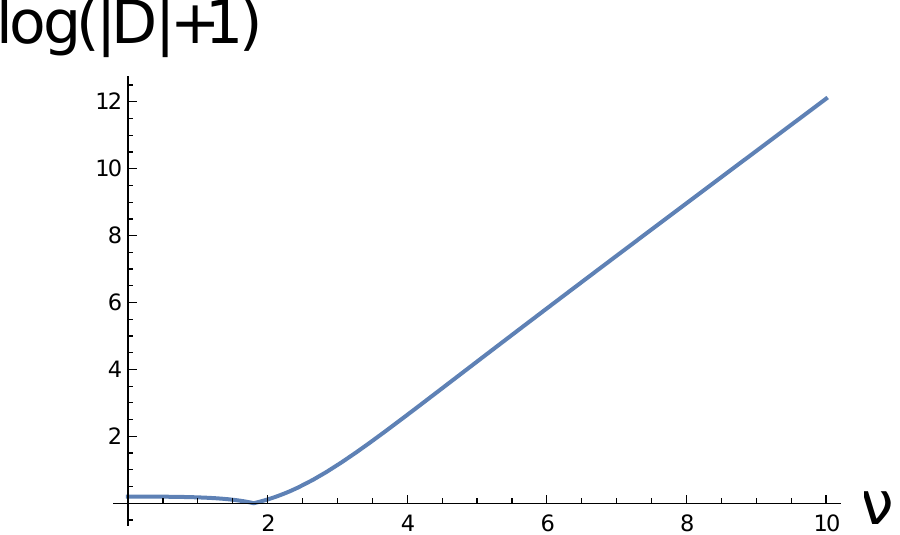}
   \includegraphics[width=7cm]{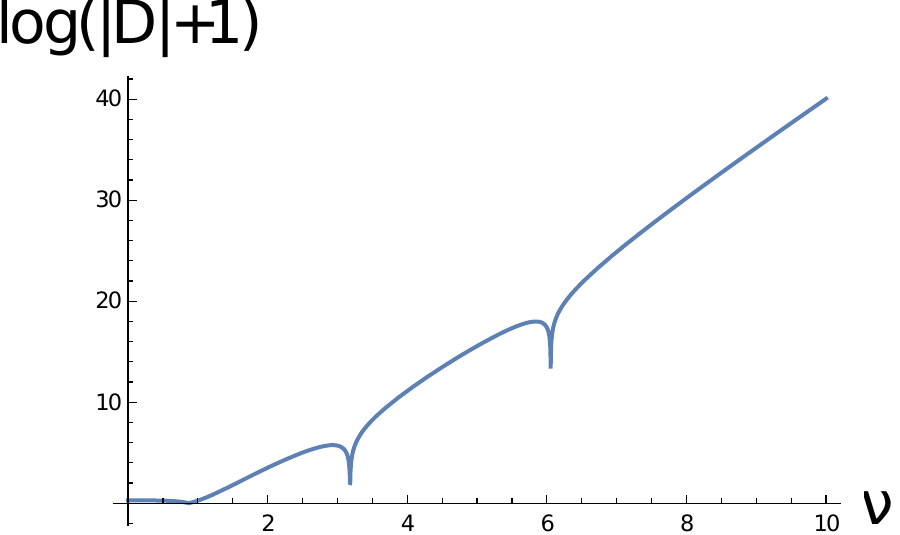}
  \caption{Plots of $\log(\abs{D(\nu, x_1,x_2)}+1)$ for $x_1=1$ and $x_2=5$ as a function of real $\nu$ (left) and plots for $x_1=0.1$ and $x_2=10$ (right). The downward pointing part of the graph indicates the zero point of $D$. }
\label{fig:ds_dir_real}
\end{figure}

\begin{figure}[hhh]
  \centering
  \includegraphics[width=7cm]{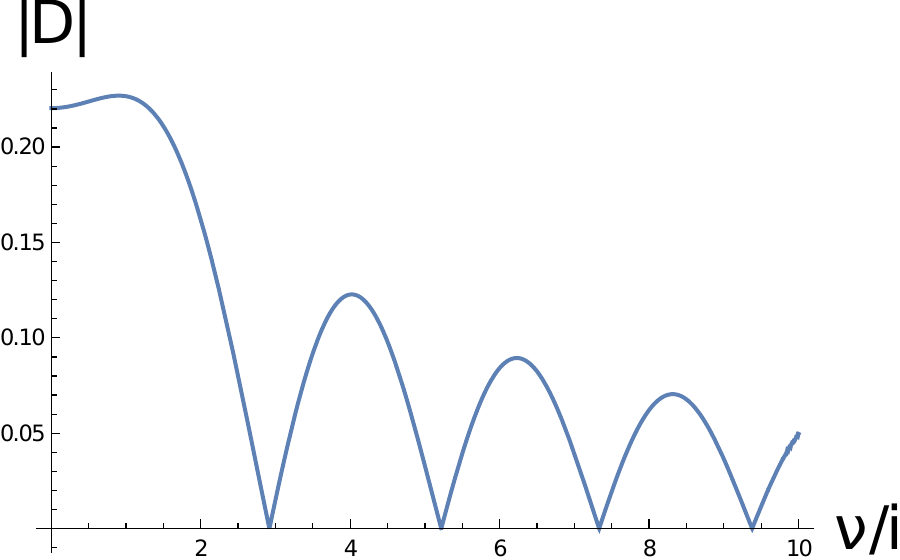}
   \includegraphics[width=7cm]{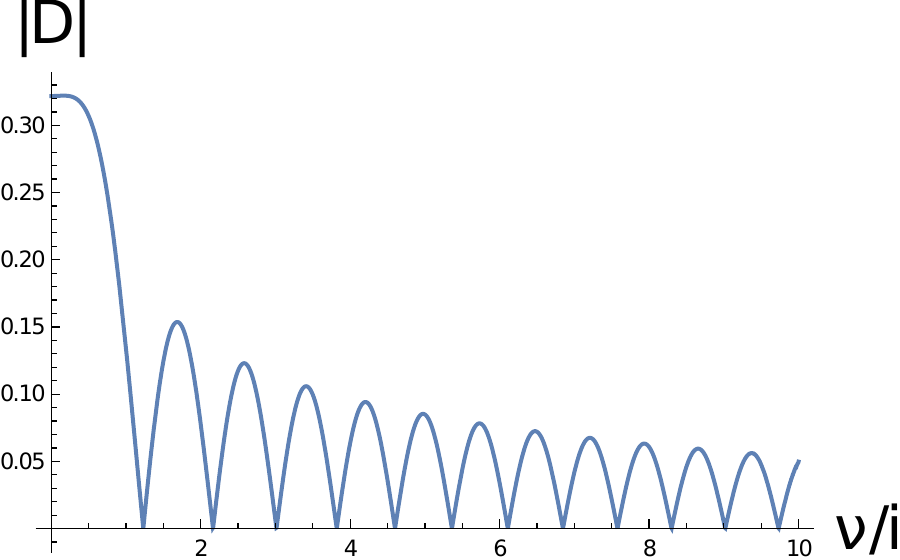}
  \caption{Plots of $\abs{D(\nu, x_1,x_2)}$ for $x_1=1$ and $x_2=5$ as a function of imaginary $\nu$ (left) and plots for $x_1=0.1$ and $x_2=10$ (right).}
\label{fig:ds_dir_im}
\end{figure}

\begin{figure}[hhh]
  \centering
  \includegraphics[width=7cm]{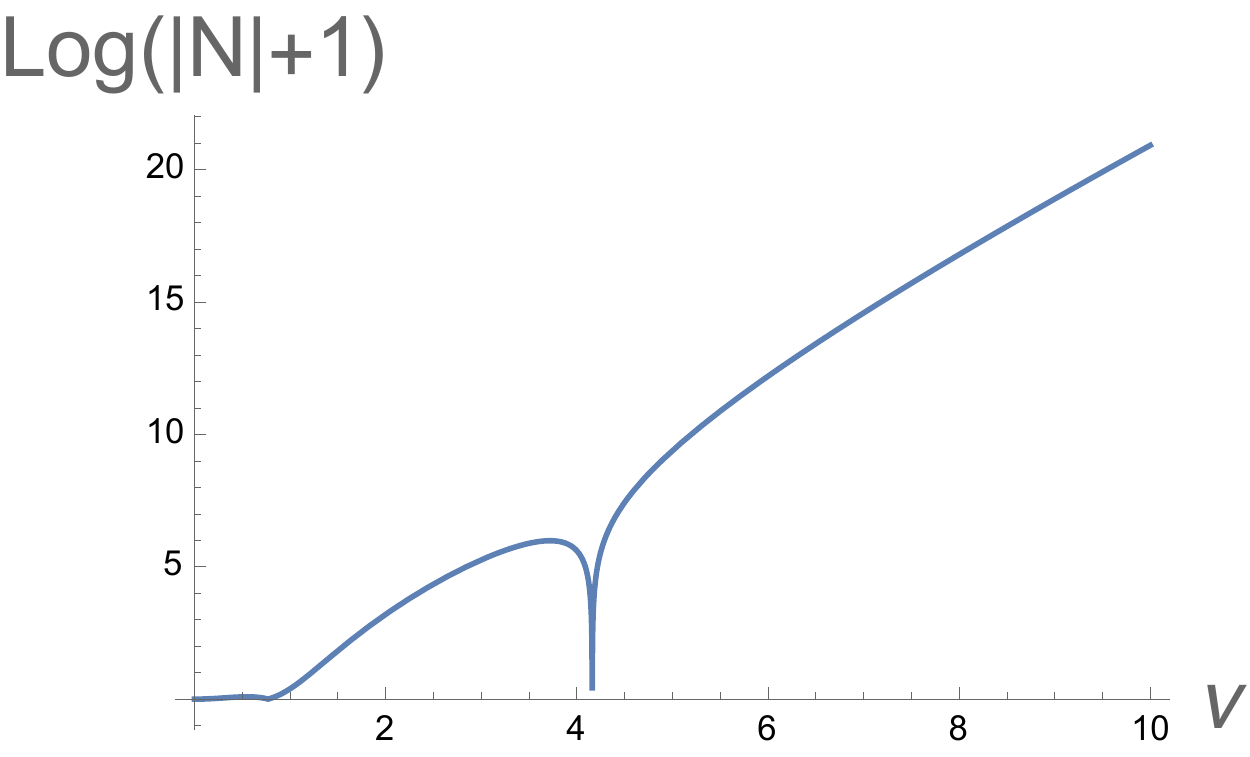}
   \includegraphics[width=7cm]{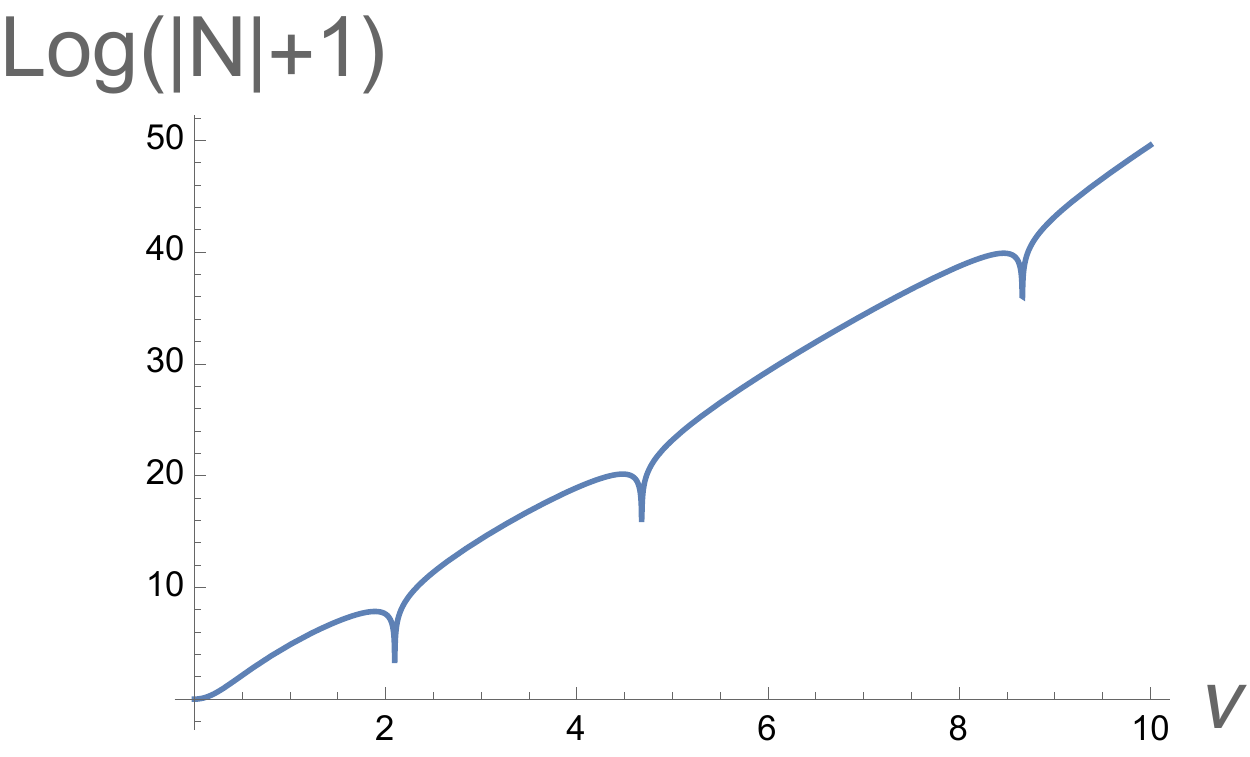}
  \caption{Plots of $\log(\abs{N(\nu, x_1,x_2)}+1)$ for $x_1=1$ and $x_2=5$as a function of real values of $\nu$ (left) and plots for $x_1=0.1$ and $x_2=10$ (right). The downward pointing part of the graph indicates the zero point of $N$. }
\label{fig:ds_neu_real}
\end{figure}

\begin{figure}[hhh]
  \centering
  \includegraphics[width=7cm]{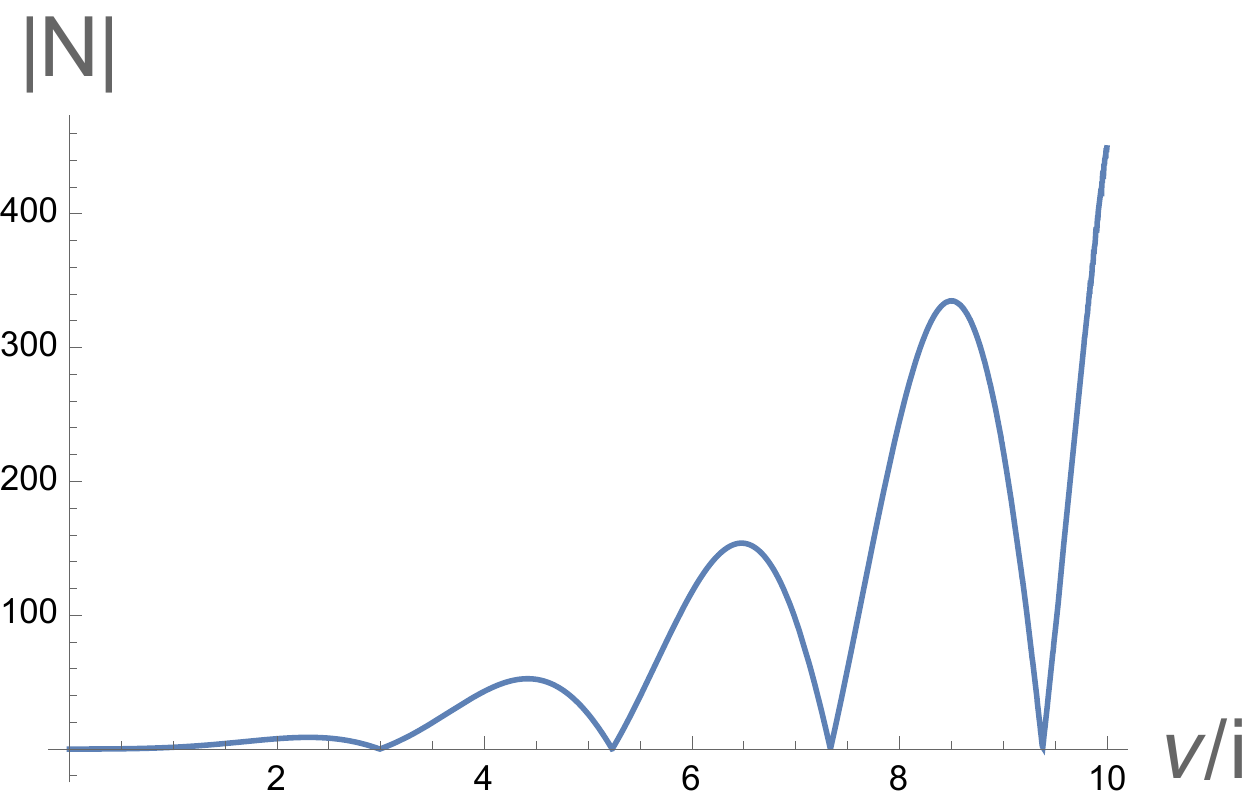}
   \includegraphics[width=7cm]{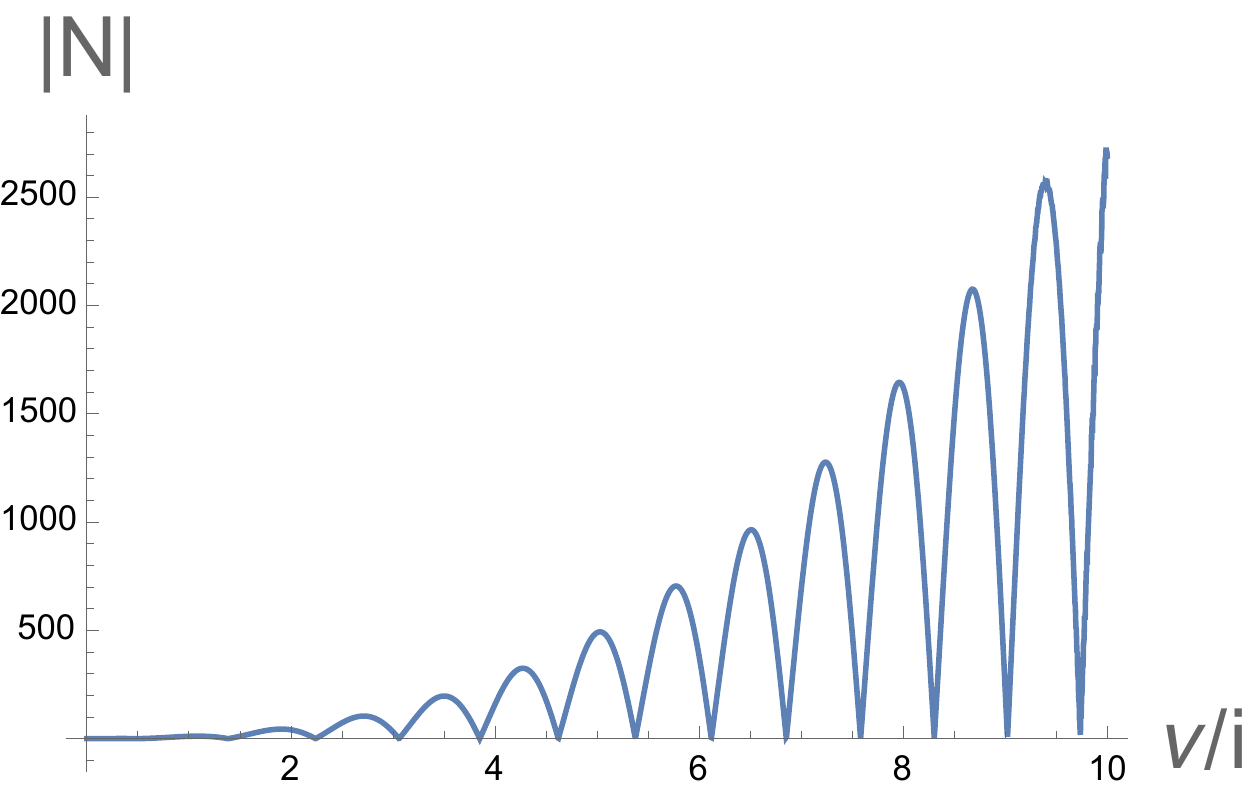}
  \caption{Plots of $N^{dS}(\nu,x_1,x_2)$ for $x_1=1$ and $x_2=5$ as a function of imaginary values of $\nu$ (left) and plots for $x_1=0.1$ and $x_2=10$ (right).  }
\label{fig:ds_neu_im}
\end{figure}

Here we present analytical calculations of scalar field modes which satisfy either Dirichlet or Neumann boundary condition in the de Sitter sliced wedges $r_1\leq r\leq r_2$. 
In Fig.\ref{fig:ds_dir_real}-\ref{fig:ds_neu_im}, $D$ and $N$ are defined in (\ref{dsDdef},\ref{dsNdef}).

\subsection{Dirichlet boundary condition}\label{sec:ApD}
As opposed to the hyperbolic slice case, 
the values of $\nu$ satisfying the boundary condition can also be real as well as pure imaginary.
We can rewrite (\ref{frti}) as following:
\be
\ti{f}(r)=\ap \frac{H_{\nu}^{(1)}(mr)}{r}+\beta\frac{H_{\nu}^{(2)}(mr)}{r}.
\ee
Then, the boundary condition (\ref{ds_dir_bndy}) can be written as
\ba
\label{ds_dir_bndy_appD}
\ti{f}(r_i)=0 &\Leftrightarrow& \ap H_{\nu}^{(1)}(x_i)+\beta H_{\nu}^{(2)}(x_i)=0 \nonumber \\
&\Leftrightarrow& \ti{\ap}=-\frac{H_{\nu}^{(2)}(x_i)}{H_{\nu}^{(1)}(x_i)}=\frac{e^{i\nu\pi} J_{\nu}(x_i)-J_{-\nu}(x_i)}{e^{-i\nu\pi} J_{\nu}(x_i)-J_{-\nu}(x_i)}
\ea
where $x_i=mr_i,i=1,2$ and $\ti{\ap}=\frac{\ap}{\beta}$.
Note that by flipping the sign of $\nu$, we obtain
\be
\ti{\ap}(-\nu)=e^{-2 \nu\pi i}\ti{\ap}(\nu),
\ee
which leads us to conclude that if $\nu$ satisfies the boundary condition (\ref{ds_dir_bndy_appD}), $-\nu$ also satisfies the condition. From the viewpoint of the numerical result, it would be sufficient to focus on only real $\nu$ and/or pure imaginary $\nu$ case. 
Before we proceed to the detailed analysis, let us review the asymptotic form of the Hankel functions. In the limit $|z|\to \infty$,
\ba
H^{(1)}_{\nu}(z)\sim\sqrt{\frac{2}{\pi z}}e^{i(z-\frac{2\nu+1}{4}\pi)},
\ \ \ H^{(2)}_{\nu}(z)\sim\sqrt{\frac{2}{\pi z}}e^{-i(z-\frac{2\nu+1}{4}\pi)}.
\ea
Also, in the region $z\sim\nu\to\infty$,
\ba
H_{\nu}{ }^{(1)}(a \nu)&=&\left\{\begin{array}{l}
-i \sqrt{\frac{2}{\pi \nu \tanh \alpha}} e^{\nu(\alpha-\tanh \alpha)}\left(1+O\left(\nu^{-1 / 5}\right)\right) ~~~~~~~~~
{[a=\operatorname{sech} \alpha<1]} \\
-\frac{1}{3 \pi} \Gamma\left(\frac{1}{3}\right)\left(e^{5 \pi i / 6}+i\right)\left(\frac{6}{\nu}\right)^{1 / 3}\left(1+O\left(\nu^{-1 / 4}\right)\right) \quad[a=1] \\
-e^{3 \pi i / 4} \sqrt{\frac{2}{\pi \nu \tan \alpha}} e^{i \nu(\tan \alpha-\alpha)}\left(1+O\left(\nu^{-1 / 5}\right)\right) ~~~~~
{[a=\sec \alpha>1]}
\end{array}\right. \\
H_{\nu}{ }^{(2)}(a \nu)&=&\left\{\begin{array}{l}
i \sqrt{\frac{2}{\pi \nu \tanh \alpha}} e^{\nu(\alpha-\tanh \alpha)}\left(1+O\left(\nu^{-1 / 5}\right)\right) ~~~~~~~~~~~~
{[a=\operatorname{sech} \alpha<1]} \\
-\frac{1}{3 \pi} \Gamma\left(\frac{1}{3}\right)\left(e^{-5 \pi i / 6}-i\right)\left(\frac{6}{\nu}\right)^{1 / 3}\left(1+O\left(\nu^{-1 / 4}\right)\right) ~~[a=1] \\
-e^{-3 \pi i / 4} \sqrt{\frac{2}{\pi \nu \tan \alpha}} e^{-i \nu(\tan \alpha-\alpha)}\left(1+O\left(\nu^{-1 / 5}\right)\right) ~~
{[a=\sec \alpha>1]}.
\end{array}\right.
\ea

Firstly, we consider the positive real $\nu$ case (remember that sign-flipped $\nu$s are also solution).
We would like to estimate $\ti{\ap}$ in $x_2\to\infty$, $x_1\to 0$. Taking $x_2$ large, we can write $\ti{\ap}$ as following:
\ba
\label{alphatildeasymptoticLarge}
\ti{\ap}&=&-\frac{H_{\nu}^{(2)}(x_2)}{H_{\nu}^{(1)}(x_2)} \no
&\sim&\left\{
\begin{array}{l}
-e^{-i\{2x_2-(\nu+\frac{1}{2})\pi \}} ~~~ (x_2>>|\nu|) \\
-1 ~~~ (x_2<|\nu|).
\end{array}
\right.
\ea
And, in small $x_1$, we can write $\ti{\ap}$ as following:
\ba
\label{alphatildeasymptoticSmall}
\ti{\ap}&=&\frac{e^{i\nu\pi} J_{\nu}(x_i)-J_{-\nu}(x_i)}{e^{-i\nu\pi} J_{\nu}(x_i)-J_{-\nu}(x_i)} \no
&\sim&\frac{e^{i\nu\pi}\gamma\left(\frac{x_1}{2}\right)^{\nu}-\left(\frac{x_1}{2}\right)^{-\nu}}{e^{-i\nu\pi}\gamma\left(\frac{x_1}{2}\right)^{\nu}-\left(\frac{x_1}{2}\right)^{-\nu}}
\ea
where $\gamma\equiv\frac{\Gamma(1-\nu)}{\Gamma(1+\nu)}$. 
When we take $\nu$ as positive real, we can solve (\ref{ds_dir_bndy_appD}) as
\ba
-e^{-i\{2x_2-(\nu+\frac{1}{2})\pi \}}&\sim& -1 ~~~ (\nu << x_2)\\
-1 &\sim& 1 ~~~ (\nu > x_2).
\ea
In the $\nu < x_2$ region, there exist solutions of $\nu$ with a period of approximately 2. Obviously, there is no solutions in the $\nu > x_2$ region. Thus, we conclude that there are finitely many solutions of real $\nu$ and the number of real solutions is bounded by $x_2$. This result is consistent with the numerical calculations, depicted in Fig.\ref{fig:ds_dir_real}.

Next, we take $\nu$ as pure imaginary $\nu=i\lambda$ and focus on positive $\lambda$ case. The conditions (\ref{alphatildeasymptoticLarge}) and (\ref{alphatildeasymptoticSmall}) are also valid even if $\nu$ is pure imaginary.
We would like to estimate $\ti{\ap}$ in $x_2\to\infty$, $x_1\to 0$. Taking $x_2$ large, we can write $\ti{\ap}$ as following:
\ba
\label{imlargeap}
\ti{\ap}&=&-\frac{H_{\nu}^{(2)}(x_2)}{H_{\nu}^{(1)}(x_2)} \no
&\sim&-e^{-i(2x_2-\frac{1}{2}\pi)} e^{-\lambda\pi}.
\ea
We can see $|\ti{\ap}|\sim e^{-\lambda\pi}$. In small $x_1$, we can write $\ti{\ap}$ as following:
\ba
\ti{\ap}&=&\frac{e^{i\nu\pi} J_{\nu}(x_i)-J_{-\nu}(x_i)}{e^{-i\nu\pi} J_{\nu}(x_i)-J_{-\nu}(x_i)} \no
&\sim&\frac{e^{-\lambda\pi}\gamma\left(\frac{x_1}{2}\right)^{i\lambda}-\left(\frac{x_1}{2}\right)^{-i\lambda}}{e^{\lambda\pi}\gamma\left(\frac{x_1}{2}\right)^{i\lambda}-\left(\frac{x_1}{2}\right)^{-i\lambda}}
\ea
Then, at large $\lambda$, we can also see $|\ti{\ap}|\sim e^{-\lambda\pi}$. Therefore, we need to focus on the phase matching of $\ti{\ap}$ in both limits.
\be
e^{\lambda\pi}\ti{\ap}=\frac{-\left(\gamma^{\frac{1}{2}}\left(\frac{x_1}{2}\right)^{i\lambda}-e^{\lambda\pi}\gamma^{-\frac{1}{2}}\left(\frac{x_1}{2}\right)^{-i\lambda}\right)^2}{|e^{\lambda\pi}\gamma \left(\frac{x_1}{2}\right)^{i\lambda}-\left(\frac{x_1}{2}\right)^{-i\lambda}|^2}
\ee
After a little calculation, we obtain
\be
\frac{\mathrm{Im}[-\ti{\ap}^{1/2}]}{\mathrm{Re}[-\ti{\ap}^{1/2}]}=-\frac{1}{\tanh \frac{\lambda\pi}{2}}\tan(\lambda\log\frac{x_1}{2}+\theta)
\ee
where $\gamma^{\frac{1}{2}}\equiv e^{i\theta}$. From (\ref{imlargeap}), 
\be
\frac{\mathrm{Im}[-\ti{\ap}^{1/2}]}{\mathrm{Re}[-\ti{\ap}^{1/2}]}=-\tan(x_1-\frac{1}{4}\pi)
\ee
We can see that infinitely many (but discrete) values of $\lambda$ yields $\tilde{f}$ satisfying the Dirichlet boundary condition. We can also see that the satisfactory values of $\lambda$ become continuous under $x_1\to0$ because $\log\frac{x_1}{2}\to-\infty$.This result is consistent with the numerical calculations, depicted in Fig.\ref{fig:ds_dir_im}.


\subsection{Neumann boundary condition}\label{sec:ApN}

From Fig.\ref{fig:ds_neu_real},we can observe the emergence of new zero points on the real axis of $\nu$ under the limit $r_2\to \infty$ and $r_1\to0$. And from Fig.\ref{fig:ds_neu_im}, we can see that the gap of each zero points of $D^{ds}$ on the imaginary axis of $\nu$ decreases as $r_2$ approaches to $\infty$ and $r_1$ to $0$. From the same calculation in Dirichlet boundary condition, we can show this numerically.

\bibliographystyle{JHEP}
\bibliography{Celestial}


\end{document}